\documentclass[preprint,showkeys,showpacs,preprintnumbers,amsmath,amssymb,epsfig]{revtex4}
\usepackage{graphicx}% Include figure files
\usepackage{dcolumn}% Align table columns on decimal point
\usepackage{bm}% bold math
\usepackage[colorlinks]{hyperref}
\begin{document}

%\preprint{APS/123-QED}

\title{A new design principle of robust onion-like networks self-organized in growth
%Prospective growth of complex networks by aggressively \\
%connecting long-distance relations
%Desired change of the vulnerability grown by 
%interwove long loops\\ in complex networks
}% Force line breaks with \\

\author{Yukio Hayashi, 
Japan Advanced Institute of Science and Technology, Ishikawa, 923-1292, Japan}
%\email{yhayashi@jaist.ac.jp}
%\affiliation{
%Japan Advanced Institute of Science and Technology,\\
%Ishikawa, 923-1292, Japan}
%\altaffiliation[Also at ]{Physics Department, XYZ University.}%Lines break automatically or can be forced with \\}
%\author{}
%\homepage{http://www.Second.institution.edu/~Charlie.Author}
%\affiliation{Hitachi Co. Ltd.,\\
%Second institution and/or address\\
%This line break forced% with \\
%}

\date{\today}% It is always \today, today,
             %  but any date may be explicitly specified

\begin{abstract}
Today's economy, production activity, and our life are sustained 
by social and technological network infrastructures, 
while new threats of network attacks by destructing loops 
have been found recently in network science.
We inversely take into account the weakness, 
and propose a new design principle for incrementally growing 
robust networks. 
The networks are self-organized by enhancing interwoven long loops. 
In particular, we consider the range-limited approximation of 
linking by intermediations in a few hops, and show the strong 
robustness in the growth without degrading efficiency of paths. 
Moreover, we demonstrate that 
the tolerance of connectivity is reformable even from extremely 
vulnerable real networks according to our proposed growing process 
with some investment. 
These results may indicate a prospective direction to 
the future growth of our network infrastructures. 
\end{abstract}

\pacs{05.65.+b, 89.75.Fb, 89.20.-a, 05.10.-a}
%05.65.+b: Self-organized systems
%89.75.Fb: Structures and organization in complex systems
%05.10.-a: Computational methods in statistical physics and nonlinear dynamics
%89.20.-a Interdisciplinary applications of physics

% Systems obeying scaling law, Markov process, 
% Numerical simulation;solution of equations,
% Random walks and Levy flights, Self-organized systems
%05.45.Df Fractals
%02.60.-x Numerical approximation and analysis
%
%\pacs{Valid PACS appear here}% PACS, the Physics and Astronomy
                             % Classification Scheme.
%\keywords{Suggested keywords}%Use showkeys class option if keyword
                              %display desired

\keywords{coexistence of efficiency and robustness, 
onion-like structure, 
long-distance relations, 
interwoven loops, 
unselfish self-organization}

\maketitle

%\section{\label{sec:level1}First-level heading:\protect\\ The line
%break was forced \lowercase{via} \textbackslash\textbackslash}
%\subsection{\label{sec:level2}Second-level heading: Formatting}

\section{Introduction}
Social and technological networks for communication, 
collaboration, trading, travel, or supply chain 
become more and more important, 
since their systems support our daily life and economy.
The connections between nodes facilitate information deliveries, 
physical logistics, and energy supplies.
Moreover, through some intermediations, the connections 
sometimes lead to 
new business chances, acquaintanceship, or 
remote control of the infrastructures efficiently. 
Some case studies in organization theory: 
the rapid recovery of Toyota group's supply chain from a large 
fire accident of their subcontract plants \cite{Nishiguchi98,Nishiguchi07}, 
world-wide economic networks with expanding business chances 
by Wenzhou people in China \cite{Nishiguchi07}, and 
the brain circulation system known as Silicon Valley (SV) model
for developing innovational high-tech industry 
with market opportunities by immigrant engineers \cite{Saxenian07} 
has been suggested the importance of 
{\it long-distance relations} for both robustness of connectivity 
and efficiency of path in a network.
The established connections via intermediations 
probably work well for managing cross-border operations. 

On the other hand, 
many social, technological, and biological infrastructural networks 
have a common 
scale-free (SF) structure \cite{Barabasi99} generated by the selfish 
preferential attachment referred to as {\it rich-get-richer} rule in 
consciously/unconsciously considering efficiency of paths between two 
nodes connected within a few hops. 
The SF networks also have an extreme vulnerability against intentional 
attacks \cite {Barabasi00}. 
However, in these several years by percolation analyses, 
it has been clarified that 
onion-like topological structure with positive degree-degree 
correlations gives the optimal robustness even for the attacks in SF 
networks \cite{Herrmann11,Tanizawa12}. 
Based on a natural but unselfish rule, 
onion-like networks can be incrementally grown by applying cooperative 
partial copying and adding shortcut \cite{Hayashi14,Hayashi16a} 
instead of the expensive whole rewiring \cite{Holme11} or 
hierarchically expanding outer ring \cite{Herrmann15}
for enhancing the positive degree-degree correlations. 
One of the drawback is that 
the robustness is weak in early stage of the growth \cite{Hayashi16a}.
While none of incremental generation of networks
has been so far based on interwoven loops, 
new threats of network attacks by destructing loops 
have been found recently \cite{Makse15,Zhou16}.
They give severer damage than the conventional intentional 
attacks \cite{Barabasi00}, and can be easily performed.
One is Collective Influence (CI) attack \cite{Makse15} 
considered for a global optimization 
to identifying the most influence nodes called influencers 
in information spreading.
Another is Belief Propagation (BP) attack \cite{Zhou16} 
derived from a message-passing approximation algorithm 
rooted by the spin glass model in statistical physics 
for the Feedback Vertex Set (FVS) problem in belonging 
to NP-hard \cite{Karp72,Kempe03}. 

Inversely taking into account the weakness caused 
by the CI and BP attacks, 
we propose a new design principle for generating robust onion-like 
networks in focusing on enhancing of long loops, whose key factor is 
long-distance relation inspired from the 
organization theory \cite{Nishiguchi07}.
Furthermore, we consider a practical approximation of the 
network generation with moderately long loops, which is based on 
range-limited intermediations for finding linked nodes 
without large costs or efforts in the growth.

\section{Self-organized growing network by a pair of attachments}
We propose a self-organized 
growing network by enhancing long loops.
After explaining the basic model, 
we consider the realistic range-limited approximation in 2.1. 
We 
estimate the degree distribution in our proposed network in 2.2. 

\subsection{Basic model and the practical approximation 
%to reduce connection cost
}
\begin{figure}[htb]
  \includegraphics[width=\textwidth]{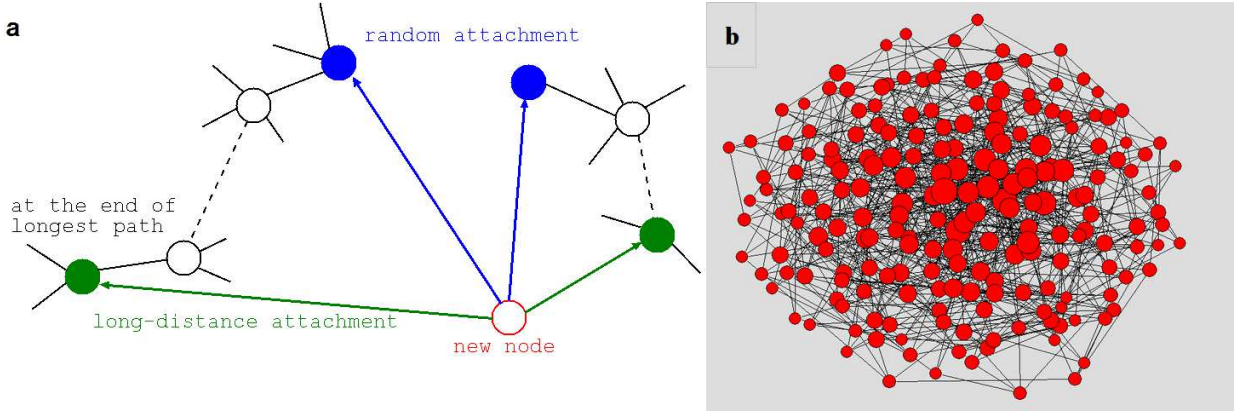}
\caption{Topological properties of the proposed networks. 
a) In the case of $m=4$, there are two pairs of 
attachments represented by green and blue lines
from a new node added at each time step. 
The green node is at the end of the longest path represented by dashed-line 
in the shortest paths counted by hops from the blue node. 
The furthest node is easily findable by a labeling method. 
Once a link is generated, it is undirected.
In PLD-A, the destination node of blue line is chosen with a probability 
proportional to its degree, 
instead of random selection in RLD-A. 
In MED, the destination node of green line is chosen in the 
$\mu+1$-th neighbors from the blue node, 
instead of the furthest node in RLD-A.
b) Example of onion-like structure by RLD-A for $m=4$ at $N=200$. 
The circle size of node is proportional to its degree. 
The structure is visualized at the positions 
as node degrees become smaller from core to peripheral.}
\label{fig1}
\end{figure}

We explain a basic model of self-organized growing network 
by enhancing long loops. 
At each time step of growing, 
a new node is added and connects to existing nodes.
As the connection rule for even number $m$ links emanated from 
the new node, 
we introduce a pair of attachments referred to as 
random and long distance attachments (RLD-A) or preferential and 
long distance attachments (PLD-A). 
The difference of the connection rule from that in the well-known 
Barab\'{a}si-Albert (BA) model \cite{Barabasi99} 
is a pair of attachments with long distance attachment.   　
As shown in Fig. \ref{fig1}a, the following pair of attachments is repeated in 
$m/2$ times at each time step.
\begin{description}
  \item[RLD-A:] One of link destination is uniformly randomly 
    chosen as encountering, and another link destination is the 
    furthest node from the chosen node. 
    When there are several candidates of the furthest with 
    a same distance counted by hops, 
    one of them is randomly selected. 
    Some kind of randomness is useful to avoid fixed weak-points 
    in the growth. 
  \item[PLD-A:] For the comparison with RLD-A, 
    instead of uniformly random selection, 
    one of the pair is preferentially chosen node 
    with a probability proportional to its degree \cite{Barabasi99}. 
    Another link destination is the furthest node from 
    the preferentially chosen node. 
\end{description}
For attached even number $m$ links, 
$m/2$ loops through the pair of nodes are created at each time step. 
The interwoven loops via new node 
are significant for $m \geq 4$ as shown in Sections 3 and 4. 
The minimum $m=4$ is corresponded to the least effort of attachment 
linking to be strongly robust network in our growing method. 
Such connection rule in Fig. \ref{fig1}a 
was not noticed because of the lack of emphasis on loops, but 
the importance of the part of long distance relation was 
covertly suggested in organization theory \cite{Nishiguchi07}.

Moreover, since range-limited approach is useful for efficiently 
investigating global property of network such as 
influencer \cite{Makse15}, centrality \cite{Toroczkai12}, or 
random percolation \cite{Radicchi16}, 
we apply it for generating robust networks. 
We consider a range-limited approximation of RLD-A 
as random and intermediated attachments (MED).
\begin{description}
  \item[MED:] Instead of the furthest node, 
    we select a distant node to the extent of a few hops 
    via intermediations from the randomly chosen pair node. 
    Intermediations in one hop mean attachments to the 2nd neighbors 
    of the randomly chosen node, 
    intermediations in two hops mean ones to the 3rd neighbors, 
    and intermediations in $\mu$ hops mean ones 
    to the $\mu + 1$-th neighbors. 
    When $\mu$ is small, the attachments to a few hops-th neighbors 
    have reality without large connection costs or efforts.
\end{description}
If a same destination node in RLD-A, PLD-A, or MED is chosen, 
other selection of pair is tried due to the prohibition of 
multiple links between two nodes.

\begin{figure}[htb]
\begin{minipage}{.495\textwidth}
  \begin{center} a) $p(k)$ by RLD-A \end{center}
  \includegraphics[width=\textwidth]{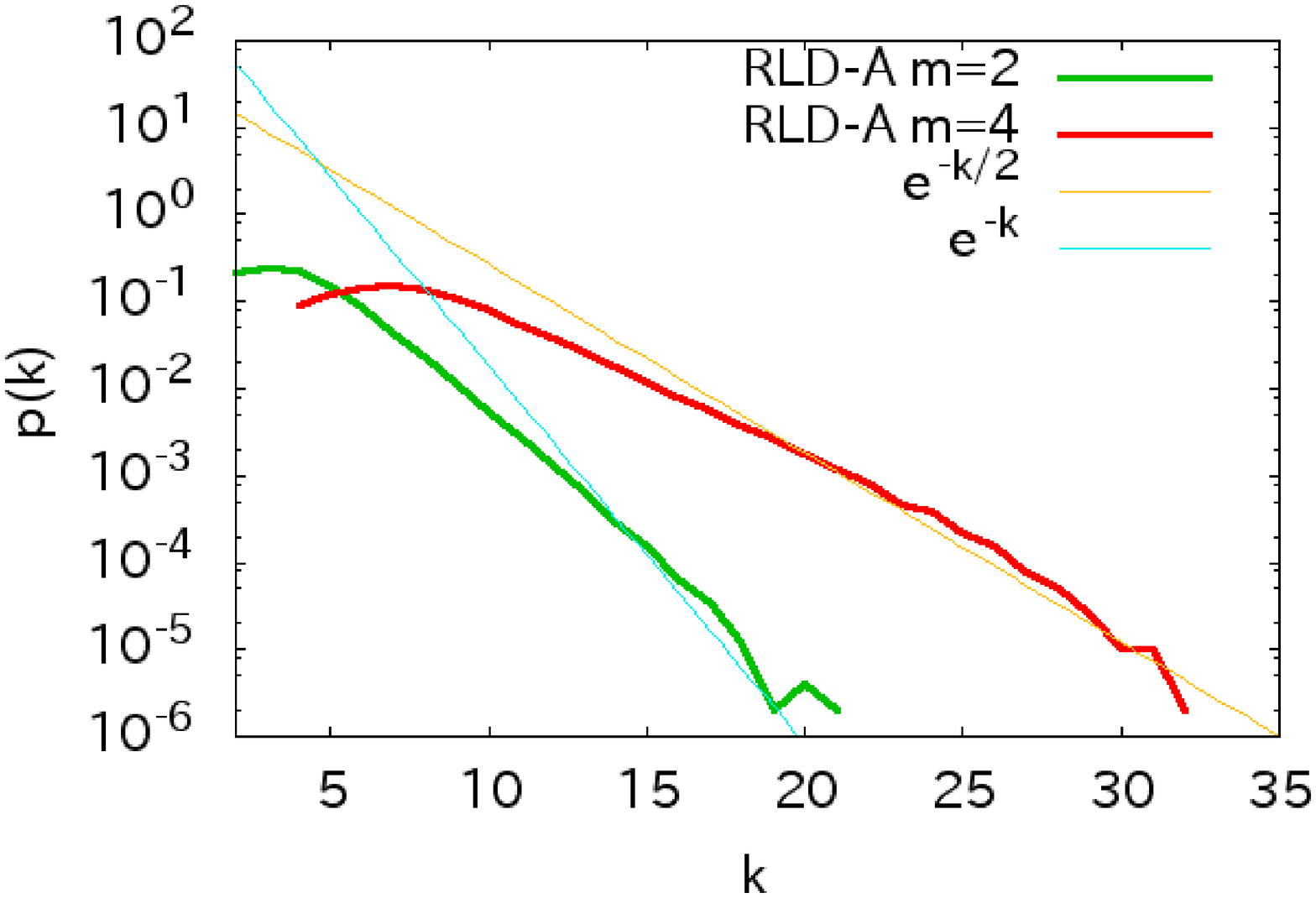}
\end{minipage}
\hfill
\begin{minipage}{.495\textwidth}
  \begin{center} b) $p(k)$ by MED \end{center}
  \includegraphics[width=\textwidth]{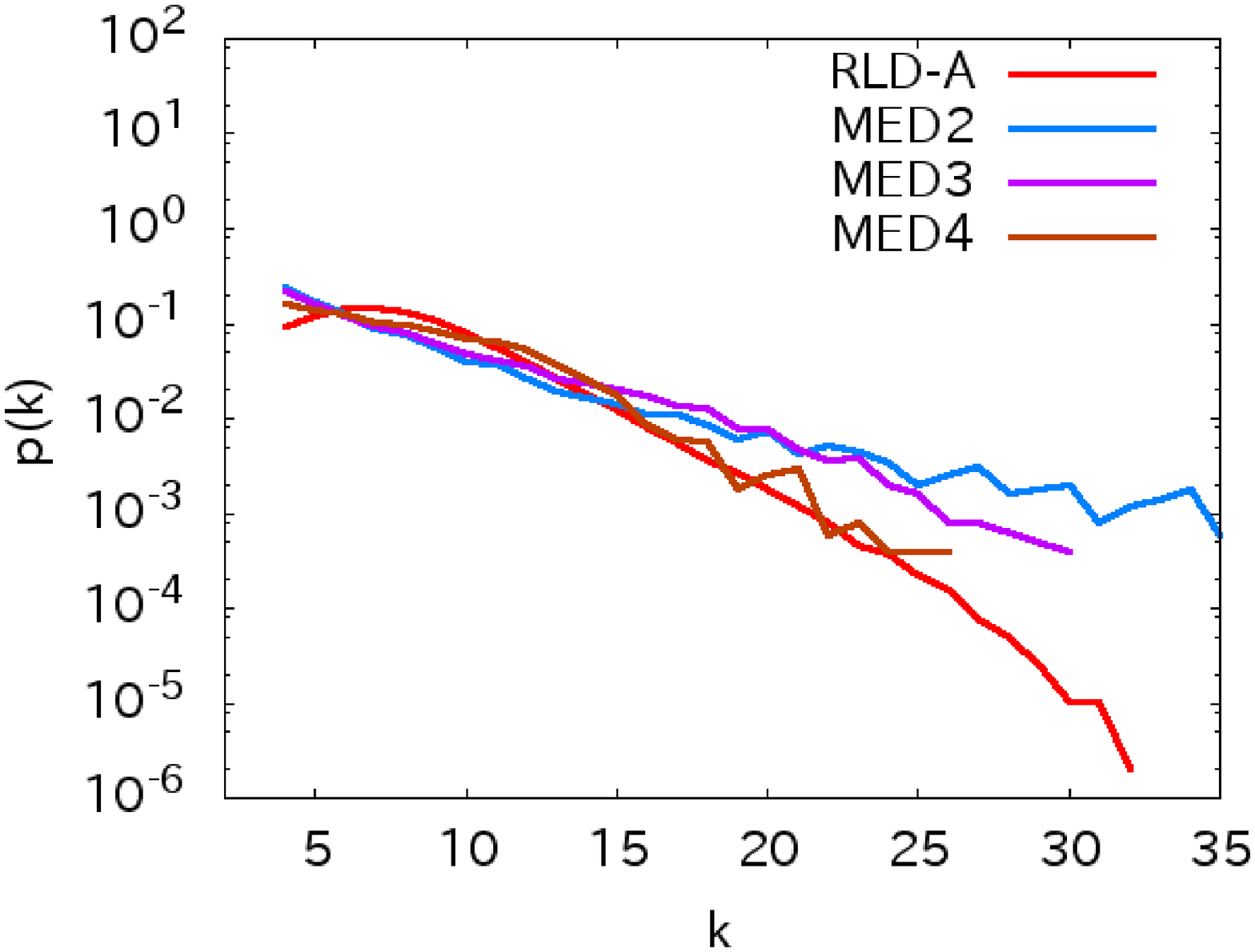}
\end{minipage}
\hfill
\begin{minipage}{.495\textwidth}
  \begin{center} c) $k_{i}(t)$ by RLD-A \end{center}
  \includegraphics[width=\textwidth]{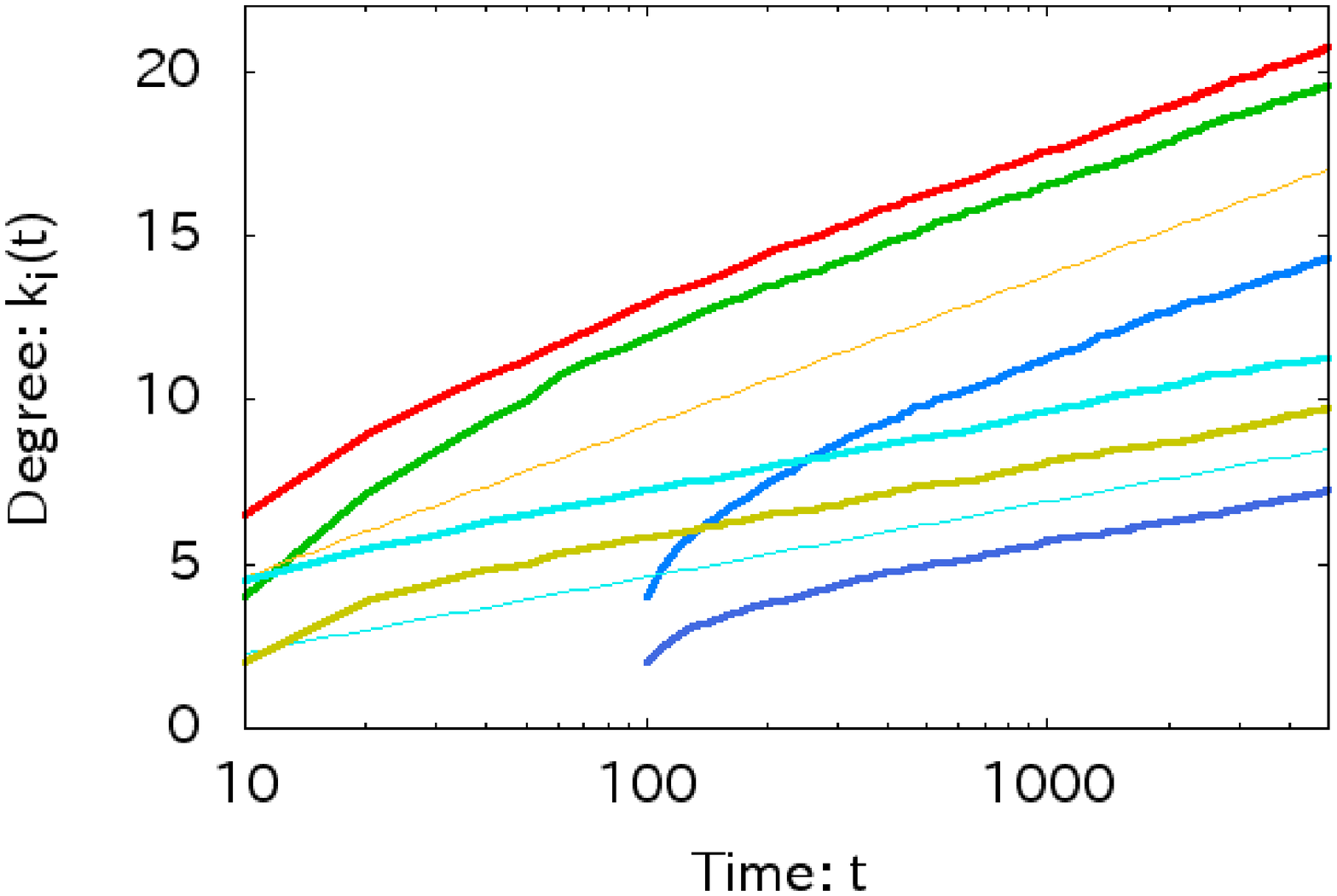}
\end{minipage}
\caption{Estimation of exponential tail of degree distribution. 
a) Degree distribution $p(k)$ in the average over $100$ samples of 
our networks at $N=5000$. 
Thin orange and cyan lines guide the exponential tails. 
b) the cases of MED for $\mu = 2, 3, 4$. 
c) Time course of degree $k_{i}(t)$ of node $i$ 
in the average over $100$ samples of our networks. 
The thick lines from top to bottom (red, green, light blue for $m=4$, 
or cyan, yellow, blue for $m=2$) denote $k_{i}(t)$ 
of node $i = 1, 10,$ and $100$ inserted at 
the birth times $t_{i} = i-m > 0$. 
Thin orange and cyan lines guide $O(\log(t))$.}
\label{fig2}
\end{figure}

\subsection{Estimation of degree distribution}
We consider growing networks with a same condition of the total 
number $M = m(N-m) + m(m-1)/2$ of links 
for size $N$: total number of nodes at time step $t = N - m$. 
As the initial configuration, 
we set a complete graph of $N_{0} = m$ nodes and 
$M_{0} = m(m-1)/2$ links at $t=0$. 
Figure \ref{fig1}b shows onion-like structure in which 
older nodes form the core while younger nodes surround it. 
Figure \ref{fig2}c justifies that older nodes get more links.
Moreover, 
we can derive exponential tails of degree distributions $p(k)$ 
by the asymptotic approximation \cite{Hayashi16b} as follows. 
The invariant ordering $k_{n}(t) < k_{n-1}(t) < \ldots < k_{1}(t)$ 
hold for large $t$ in parallel curves in Fig. \ref{fig2}c. 
Since the time course of degree of node $i$ follows 
$k_{i}(t) \sim \log(t) / \beta$ as a monotone increasing function of 
$t$ with a constant $\beta > 0$, 
we obtain 
\[
  \frac{t_{i}}{t} = \frac{e^{\beta k_{i}}}{e^{\beta k_{i}(t)}}, 
\]
\[
  p(k_{i}(t) < k) = 
  p\left(t_{i} > \frac{e^{\beta k_{i}}}{e^{\beta k}} t \right) 
  = \left( 1 - \frac{e^{\beta k_{i}}}{e^{\beta k}} \right) 
  \frac{t}{N_{0} + t},
\]
\[
  p(k) = \frac{\partial p((k_{i}(t) < k)}{\partial k} 
  \sim e^{- \beta k}.
\]
Indeed, the orange and cyan lines guide $\log(t)$ and $2 \log(t)$ 
in Fig. \ref{fig2}c, the estimated same color lines of 
$e^{-k}$ and $e^{-k/2}$ are fitting with the tails of $p(k)$ 
in Fig. \ref{fig2}a.
The largest degree is bounded around $20 \sim 35$ without 
heavy connection load as on hub nodes in SF structure of 
many real networks.
Figure \ref{fig2}b shows 
that the range-limited cases of MED in $\mu = 2, 3, 4$ intermediations 
have slightly deviated but similar exponential tails of $p(k)$.

\begin{figure}[htb]
\begin{minipage}{.495\textwidth}
  \begin{center} a) RLD-A or PLD-A for $m=4$ \end{center}
  \includegraphics[width=\textwidth]{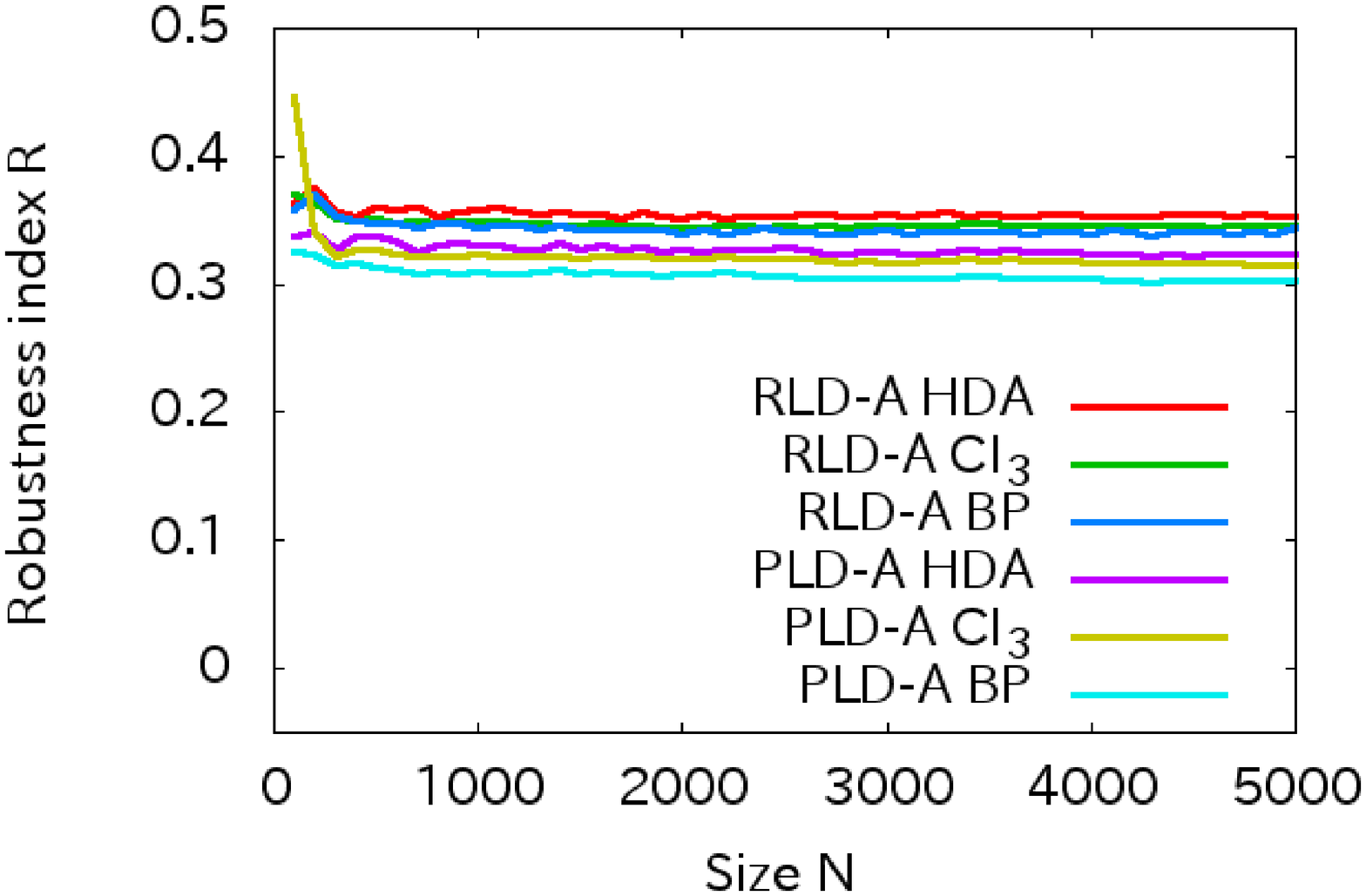}
\end{minipage}
\hfill
\begin{minipage}{.495\textwidth}
  \begin{center} b) BA for $m=4$ \end{center}
  \includegraphics[width=\textwidth]{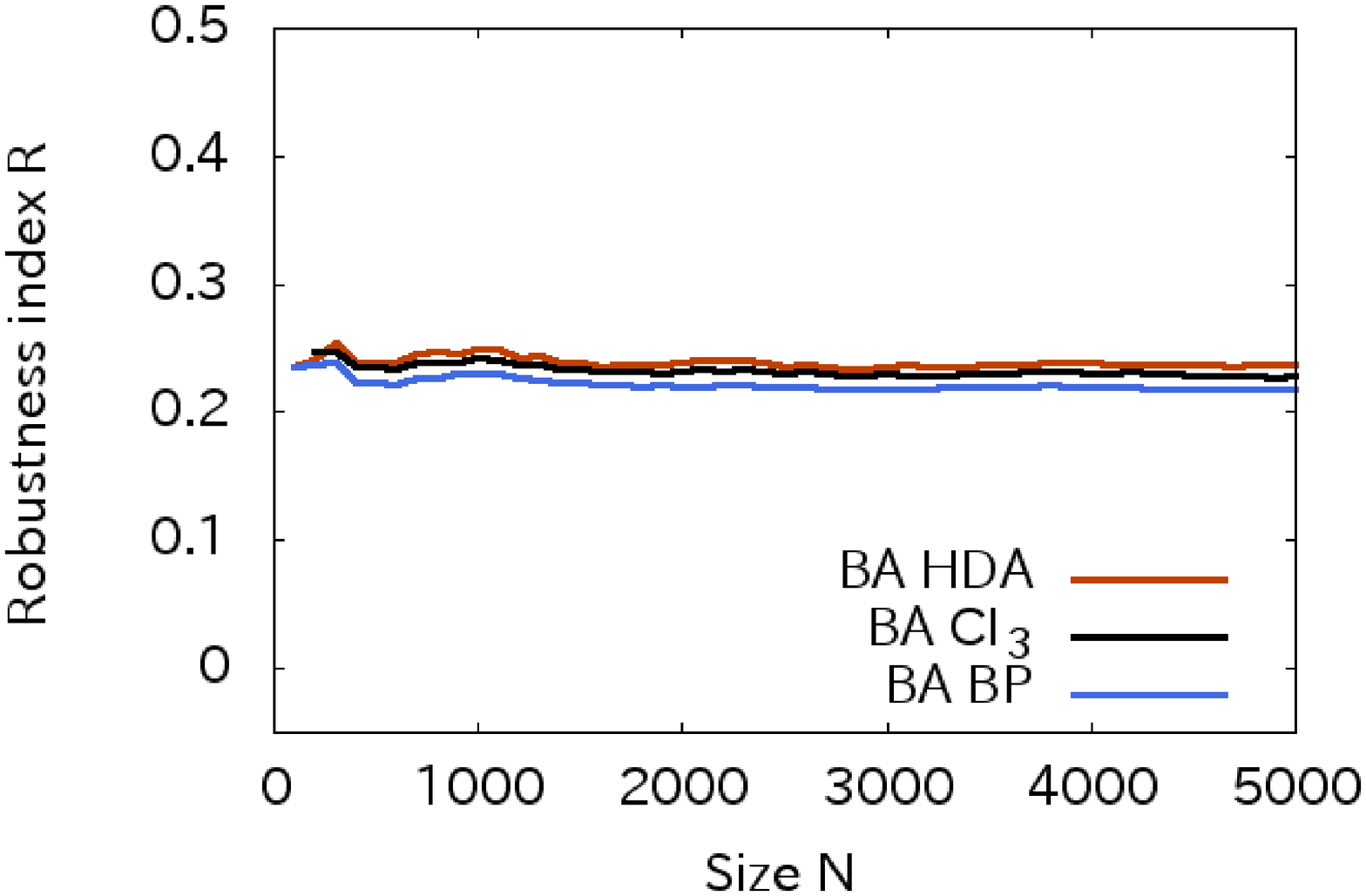}
\end{minipage}
\hfill
\begin{minipage}{.495\textwidth}
  \begin{center} c) RLD-A or PLD-A for $m=2$ \end{center}
  \includegraphics[width=\textwidth]{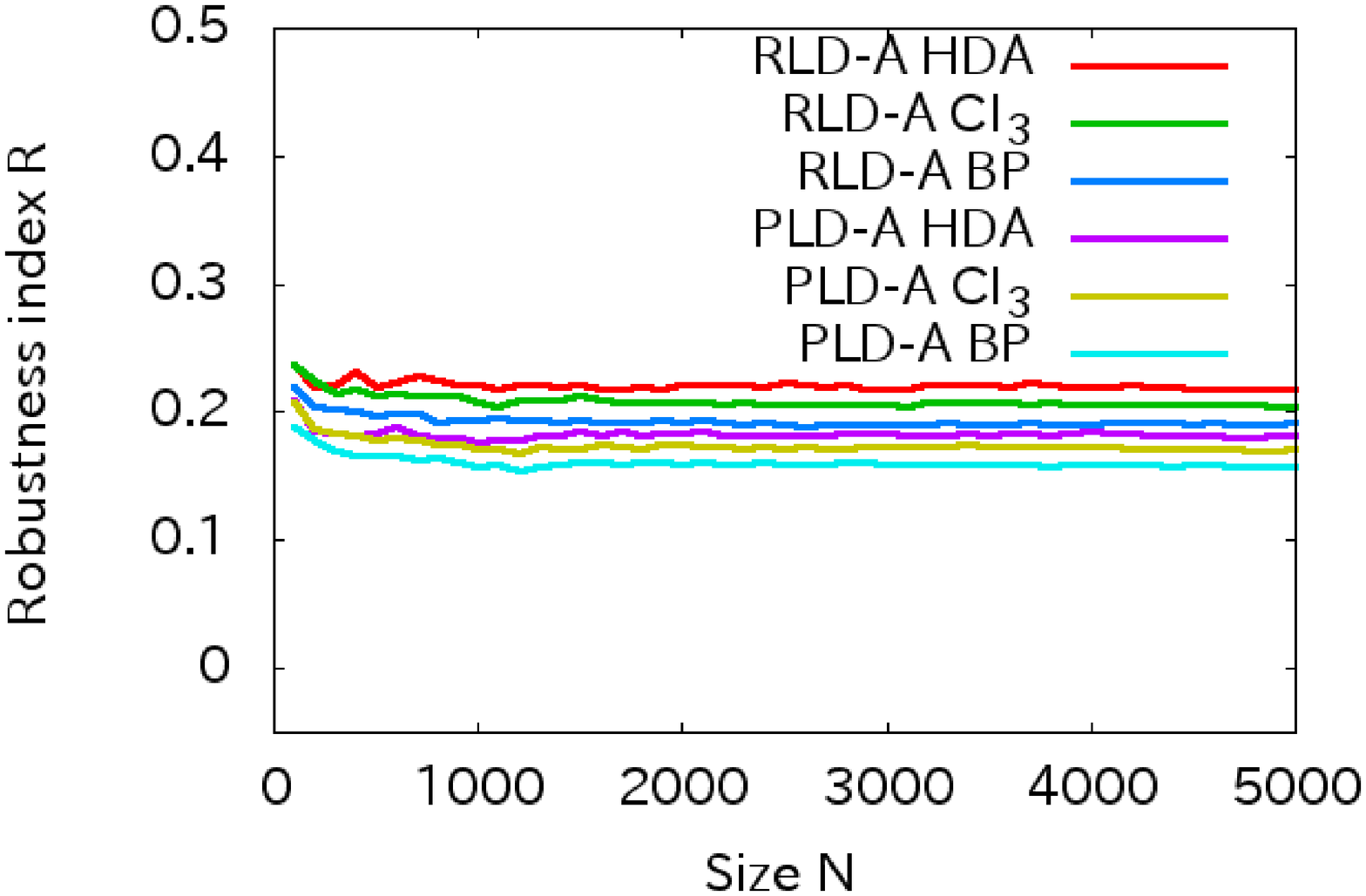}
\end{minipage}
\caption{Robustness in the growing networks. 
Robustness index $R$ against HDA, CI$_{3}$, and BP attacks 
vs size $N$ in our networks by 
a) RLD-A or PLD-A, b) SF networks by BA model for $m=4$, 
and c) RLD-A or PLD-A for $m=2$. 
}
\label{fig3}
\end{figure}

\begin{figure}[htb]
\begin{minipage}{.495\textwidth}
  \begin{center} a) RLD-A or PLD-A for $m=4$ \end{center}
  \includegraphics[width=\textwidth]{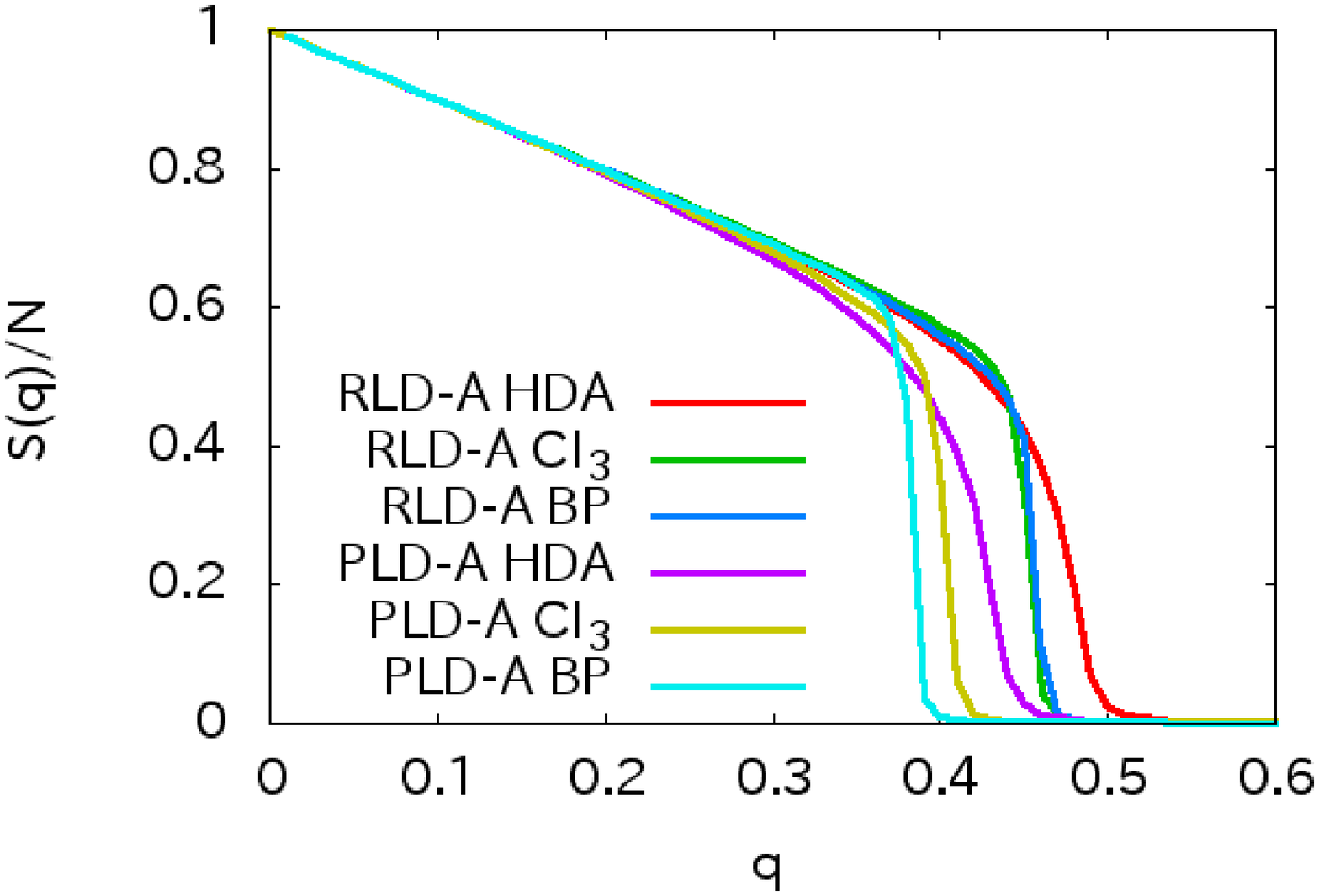}
\end{minipage}
\hfill
\begin{minipage}{.495\textwidth}
  \begin{center} b) BA for $m=4$ \end{center}
  \includegraphics[width=\textwidth]{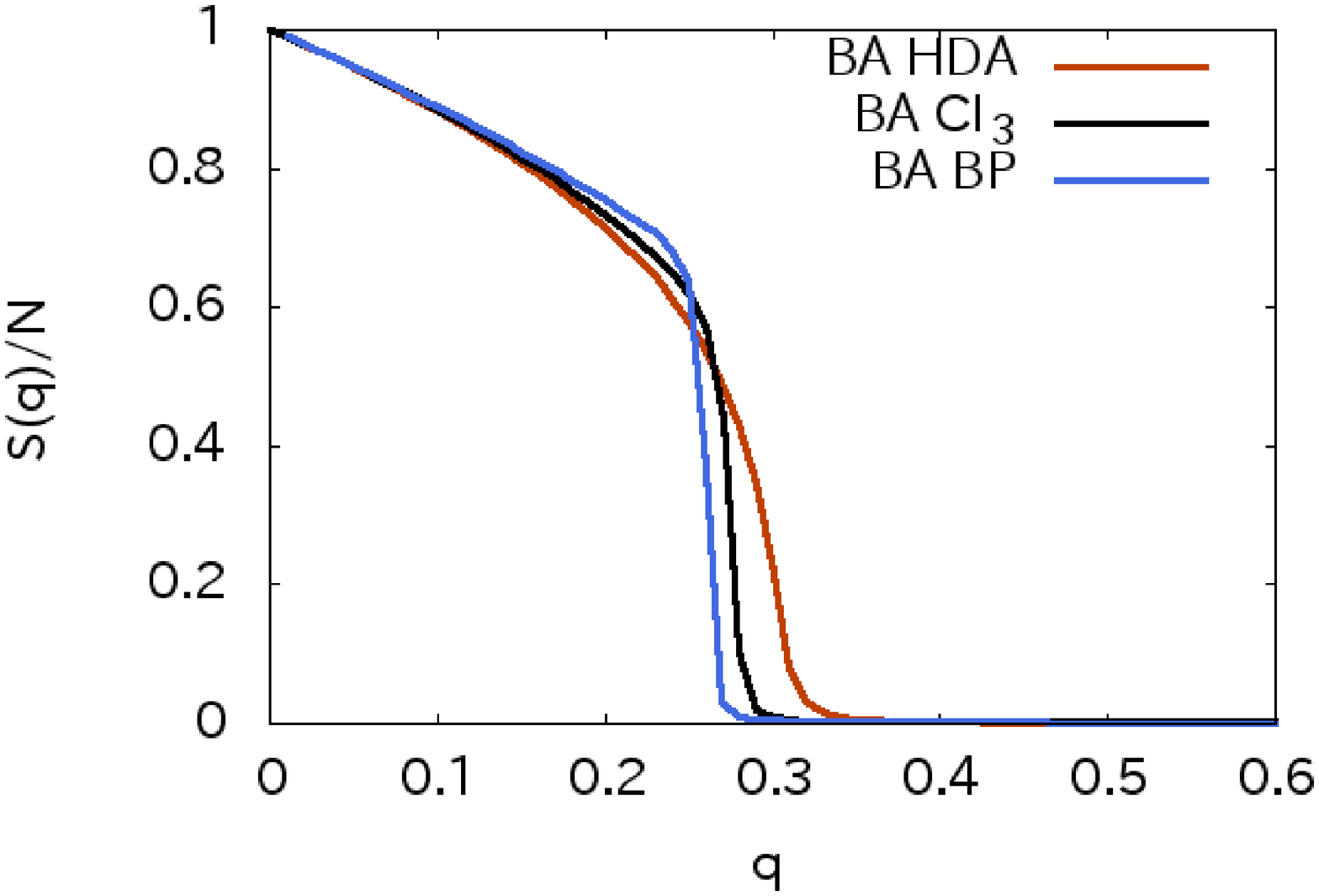}
\end{minipage}
\hfill
\begin{minipage}{.495\textwidth}
  \begin{center} c) RLD-A, PLD-A, BA for $m=2$ \end{center}
  \includegraphics[width=\textwidth]{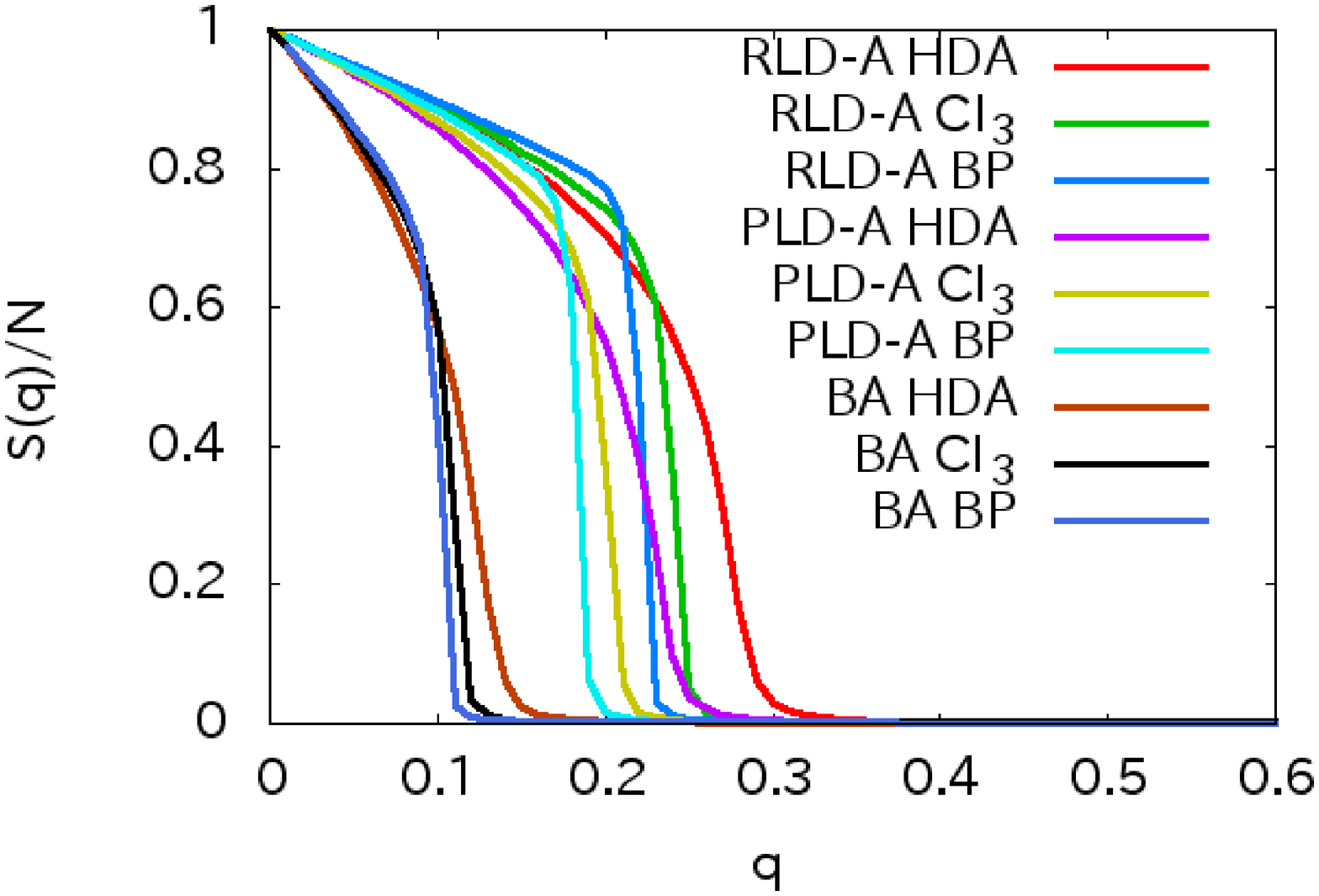}
\end{minipage}
\caption{Relative size $S(q)/N$ vs fraction $q$ of removed nodes 
in the networks by 
a) RLD-A, PLD-A, 
b) BA for $m=4$, 
and c) RLD-A, PLD-A, BA for $m=2$ at $N=5000$. 
The reddish curves against HDA attacks are gradually decreased, 
while bluish ones against BP attacks are suddenly dropped. 
The yellow-greenish or black 
curves against CI$_{3}$ attacks are the intermediate. 
Note that $R$ is defined by the area under the line of $S(q)/N$.
}
\label{fig4}
\end{figure}

\begin{figure}[htb]
\begin{minipage}{.495\textwidth}
  \begin{center} a) RLD-A, PLD-A, and BA  \end{center}
  \includegraphics[width=\textwidth]{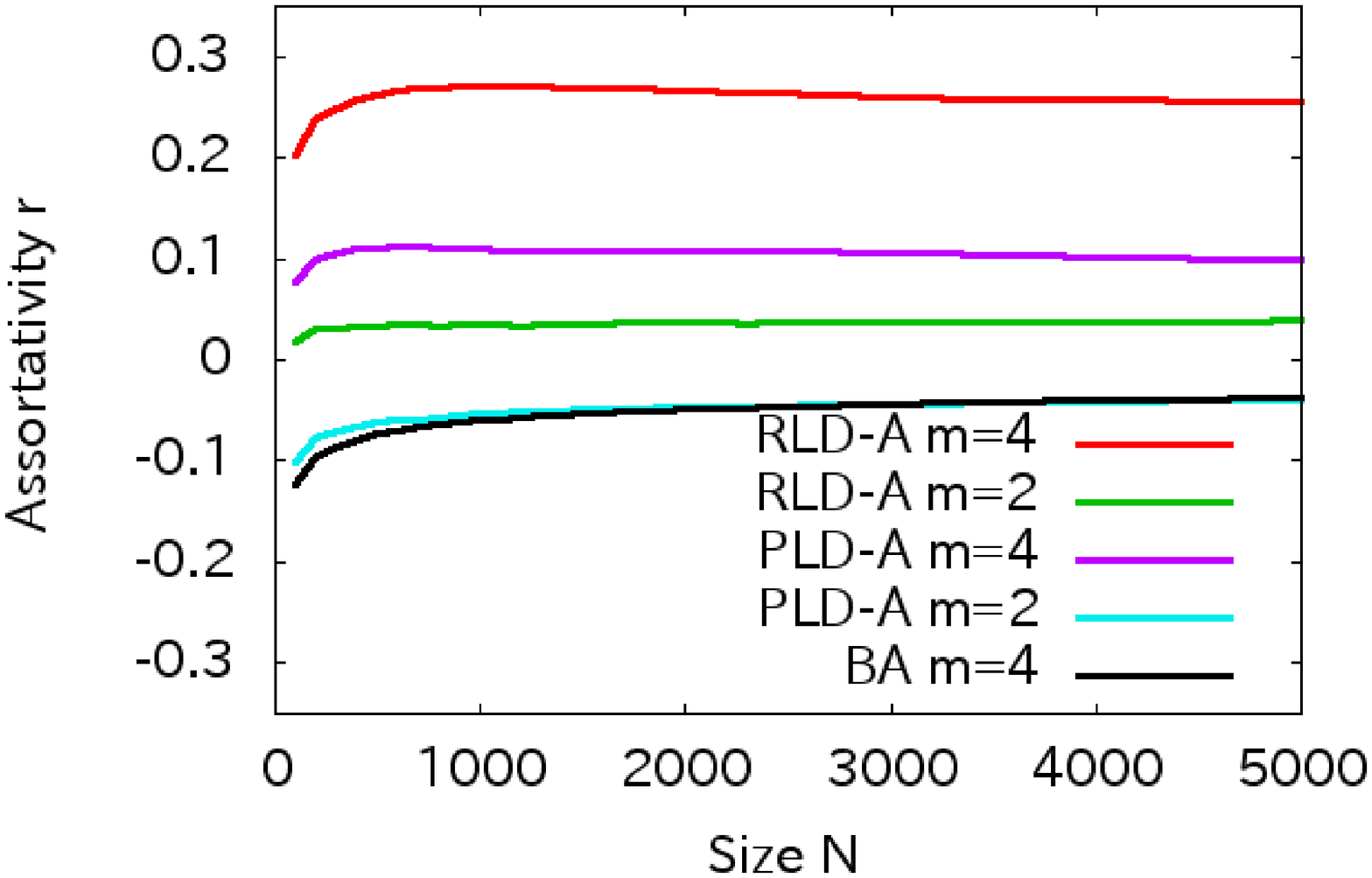}
\end{minipage}
\hfill
\begin{minipage}{.495\textwidth}
  \begin{center} b) RLD-A and MED \end{center}
  \includegraphics[width=\textwidth]{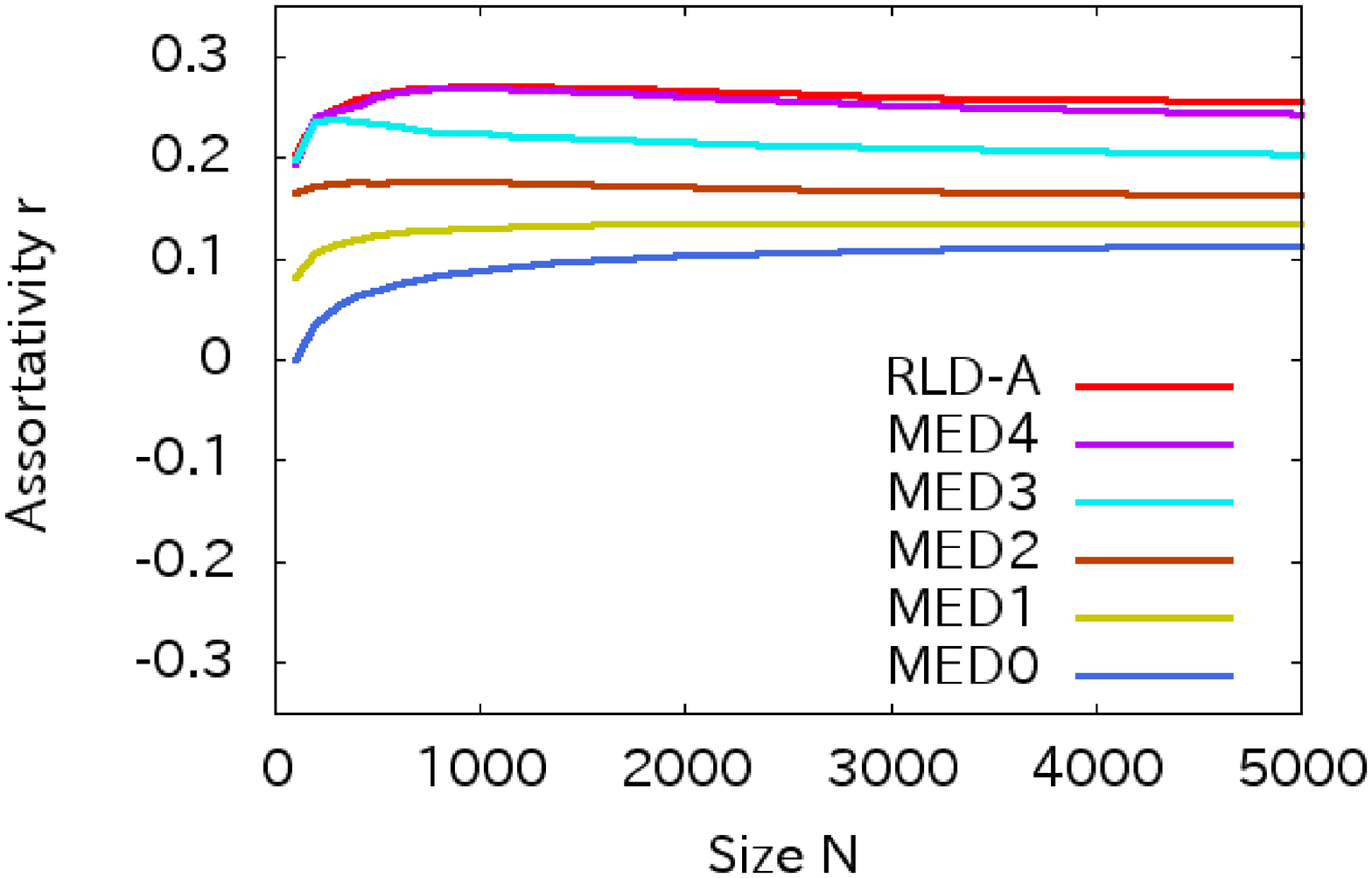}
\end{minipage}
\caption{Degree-degree correlations 
in the growing networks. 
Assortativity $r$ as the measure of correlations
for size $N$ in comparison with the networks by a) RLD-A, PLD-A, and 
BA model for $m=2$ or $m=4$, b) RLD-A and MED0$\sim$4 for $m=4$. 
%To be an onion-like network, a certain large $r > 0.2$ is necessary.
MED$0 \sim 4$ denote the case of MED for $\mu = 0, 1, 2, 3, 4$.
}
\label{fig5}
\end{figure}

\begin{figure}[htb]
  \includegraphics[width=0.62\textwidth]{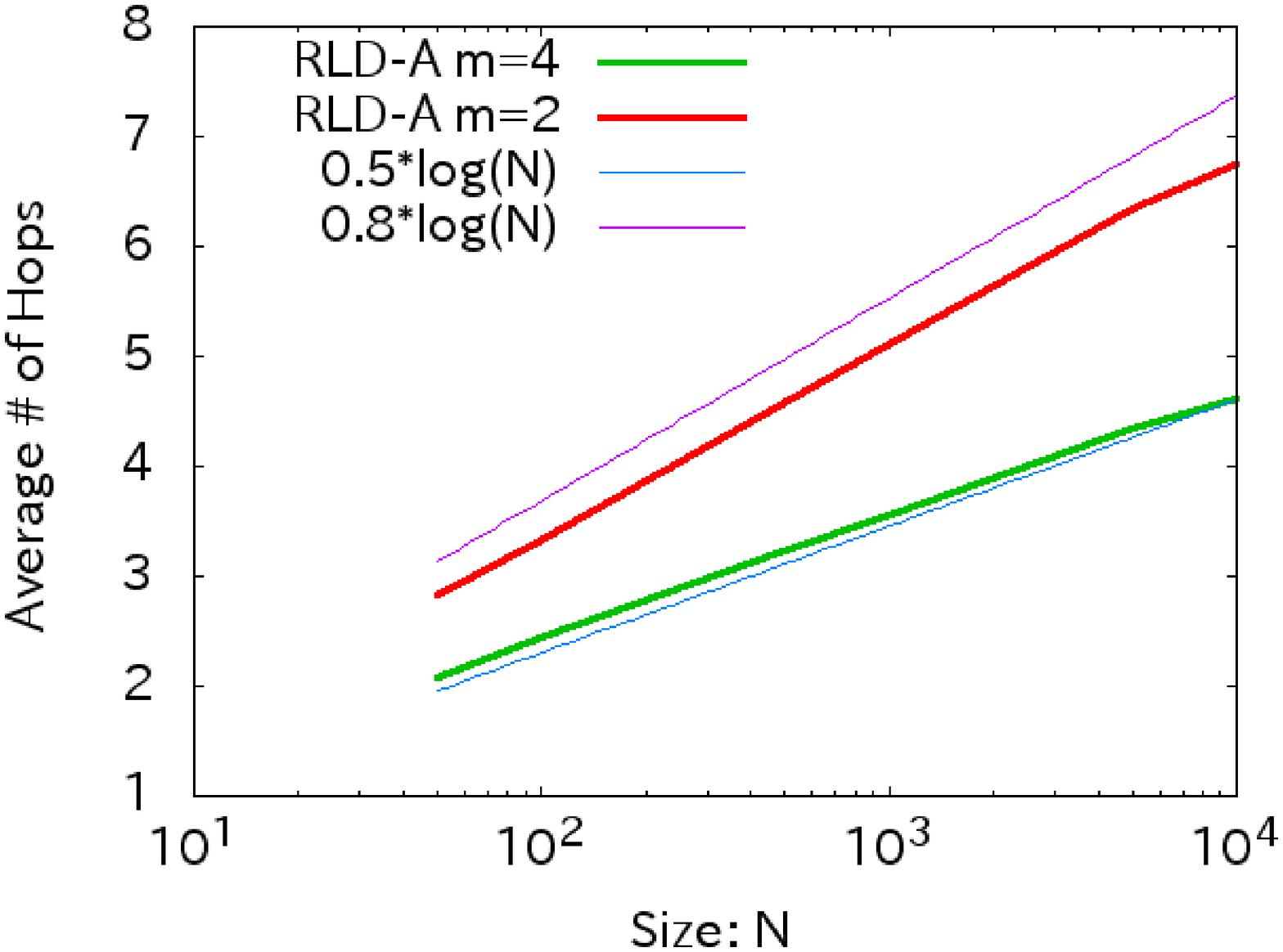}
\caption{Average path length on the shortest paths counted by hops
in our growing networks by RLD-A. 
The purple and blue thin lines guide $O(\log(N))$ as the small-world effect.
These results are averaged over $100$ samples. }
\label{fig6}
\end{figure}

\section{Strong robustness and the small-world effect}
For our proposed networks, 
we investigate the robustness index 
\[
 R \stackrel{\rm def}{=} \frac{1}{N} \sum_{q = 1/N}^{1} S(q),
\]
where $S(q)$ denotes the number of nodes included in the giant 
component (GC as the largest cluster) 
after removing $q N$ nodes, $q$ is a fraction of removed 
nodes by High Degree Adaptive (HDA), 
CI for $l=3$ layer \cite{Makse15}, 
and BP \cite{Zhou16} attacks.
As in Appendix or  \cite{Makse15,Zhou16}, 
the highest value of $CI_{l}(i)$ in Eq. (\ref{eq_CI})
or $q^{0}_{i}$ in Eqs. (\ref{eq_BP1})-(\ref{eq_BP4})
to be removed is recalculated after each node removal. 
Note that the maximum $R \geq 0$ is $0.5$ in general.
The following results are insensitive for varying values of 
inverse temperature $x=7$ and $100$ rounds 
of the message-passing \cite{Zhou16}, 
and there is no difference for $l \geq 3$ in CI attacks. 
Figure \ref{fig3}a shows that 
our networks by RLD-A for $m=4$ have strong robustness $R > 0.3$ 
even in the early stage of growth, 
while Figure \ref{fig3}b shows 
that $R$ is lower in the conventional SF networks by BA model. 
The networks by PLD-A show the intermediate $R$ values. 
In Fig. \ref{fig3}c for $m=2$, 
these lines fall in overall, but it is invariant that 
the ordering of damage by attacks is BP $>$ CI$_{3}$ $>$ HDA
whose differences are very small.
Each value of $R$ is almost constant in the growing 
at least from the initial complete graph. 
In the range-limited cases of MED in $\mu = 2, 3, 4$ intermediations, 
we obtain $0.31 < R < 0.35$ and $0.28 < R < 0.34$ against 
HDA and BP attacks, respectively. 
Figure \ref{fig4}ab show the relative size $S(q)/N$ with 
the sudden breakdowns by BP attacks (bluish lines) as 
mentioned in \cite{Zhou16}. 
Each of the robustness in Fig. \ref{fig4}ab for $m=4$ 
is improved from the corresponding one in Fig. \ref{fig4}c for 
$m=2$, although larger $m$ requires more links. 

We also investigate the assortativity $-1 \leq r \leq 1$ 
as the Pearson correlation coefficient for degrees \cite{Newman03}.
\[
  r \stackrel{\rm def}{=} \frac{4 M \sum_{e} (k_{e} k'_{e}) 
  - \left[ \sum_{e} (k_{e} + k'_{e}) \right]^{2}}{
  2 M \sum_{e} (k^{2}_{e} + k'^{2}_{e}) 
  - \left[ \sum_{e} (k_{e} + k'_{e}) \right]^{2}},
\]
where $k_{e}$ and $k'_{e}$ denote degrees at both end-nodes 
of link $e$, $M$ is the total number of links. 
Figure \ref{fig5}a shows that 
our networks by RLD-A for $m=4$ (red line) 
have high assortativity $r > 0.2$ as similar to 
the copying model \cite{Hayashi14,Hayashi16a}. 
However PLD-A is insufficient to create strong correlations. 
Figure \ref{fig5}b shows that 
the range-limited cases of MED in $\mu = 3, 4$ (purple and cyan lines) 
are close to the case of RLD-A (red line). 
Although there is no clear criteria for the value of $r$ 
in order to be an onion-like network with necessary 
positive degree-degree correlations, 
too large $r$ is unsuitable \cite{Tanizawa12}. 
We do not discuss the optimally robust onion structure, 
but concern about incrementally growing 
proper good onion-like networks self-organized by natural and 
reasonable attachments. 
From Figs. \ref{fig3} \ref{fig4} and \ref{fig5}, 
our networks by RLD-A and MED in $\mu = 3, 4$ for $m = 4$ 
have onion-like structure with both high $R$ and $r$, 
but other cases are not.
Moreover, 
they have efficient small-world property \cite{Watts98}: 
the average shortest path length is $O(\log(N))$ as shown 
in Fig. \ref{fig6}, 
even though half links in RLD-A or MED 
are created by random attachment without intention to be efficiency. 
In the growing from the initial complete graph, 
the number $\mu \approx 3$ of intermediations 
is at the similar level of the average path length.
Averagely the length of simple one-round loop 
(as shown in Fig. \ref{fig1}a, it consists of the 
path between blue and green nodes 
$+$ the corresponding blue and green links) 
generated by the pair of RLD-A or MED 
becomes short and inexpensive as $O(\log(N))$.

\begin{table}[hbt]
\caption{Basic data of the real networks 
%\cite{dataset} 
after converting from each of them to an undirected graph 
without multiple links. 
USair and USpower are abbreviations of US Airport Network and 
US Power Grid, respectively.
The average path length and diameter are defined by 
the averaged length of the shortest paths with the minimum 
number of hops between two nodes
and the longest length in a network.}\label{table1}
\begin{minipage}{\textwidth}
{\tabcolsep = 2.6mm
\begin{tabular}{cccccc} \hline\hline
 Network & $N_{0}$ &  $M_{0}$   & $\langle k \rangle = 2 M_{0} / N_{0}$ 
 & Ave. path len. & Diameter\\ \hline 
Facebook & 1899 & 13838 & 14.573 & 3.055 & 8 \\
USair    & 1574 & 17215 & 21.874 & 3.115 & 8 \\
USpower  & 4941 & 6594  & 2.669  & 18.989 & 46 \\ \hline\hline
\end{tabular} 
}
\end{minipage}
\end{table}

\section{Virtual test for prospective growth of real networks}
As a virtual test for exploring future design of networks, 
we study the robustness 
of our model in growing to onion-like structure 
from the initial configuration of 
real networks\footnote{http://toreopsahl.com/datasets/} 
in Table \ref{table1}. 
For these social and technological networks, 
long distance connections will be somewhat required in order to 
%open the door 
seek solution strategies 
to undeveloped relationship or inconvenience, 
and realizable by intermediation or investment (e.g. 
for low cost carrier or innovation of power transmission)
in a trade-off between the benefit and the cost.

\begin{figure}[htb]
\begin{minipage}{.39\textwidth}
  \begin{center} a) Facebook for $m=4$ \end{center}
  \includegraphics[width=\textwidth]{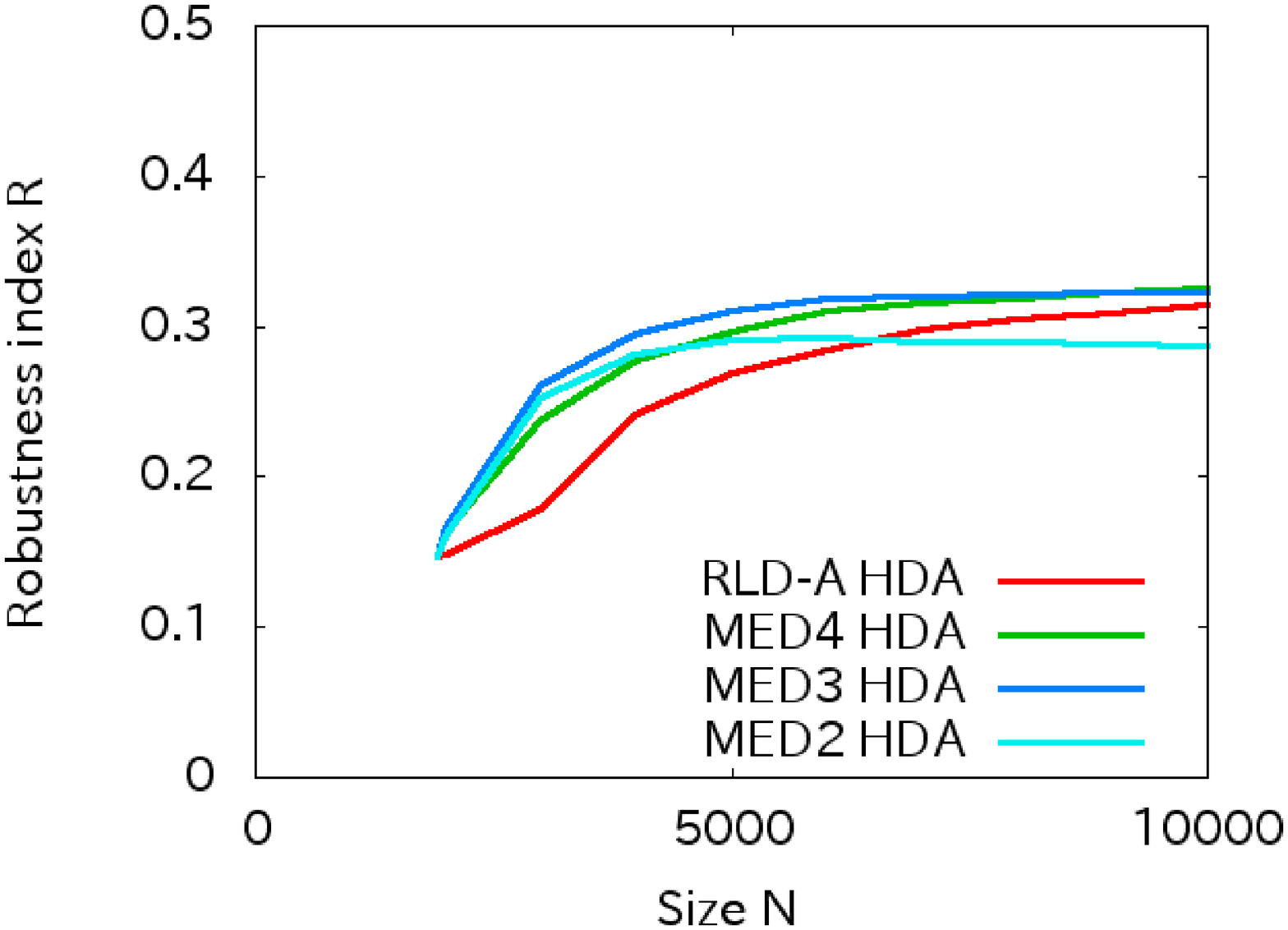}
\end{minipage}
\hfill
\begin{minipage}{.39\textwidth}
  \begin{center} b) US Airport Network for $m=4$ \end{center}
  \includegraphics[width=\textwidth]{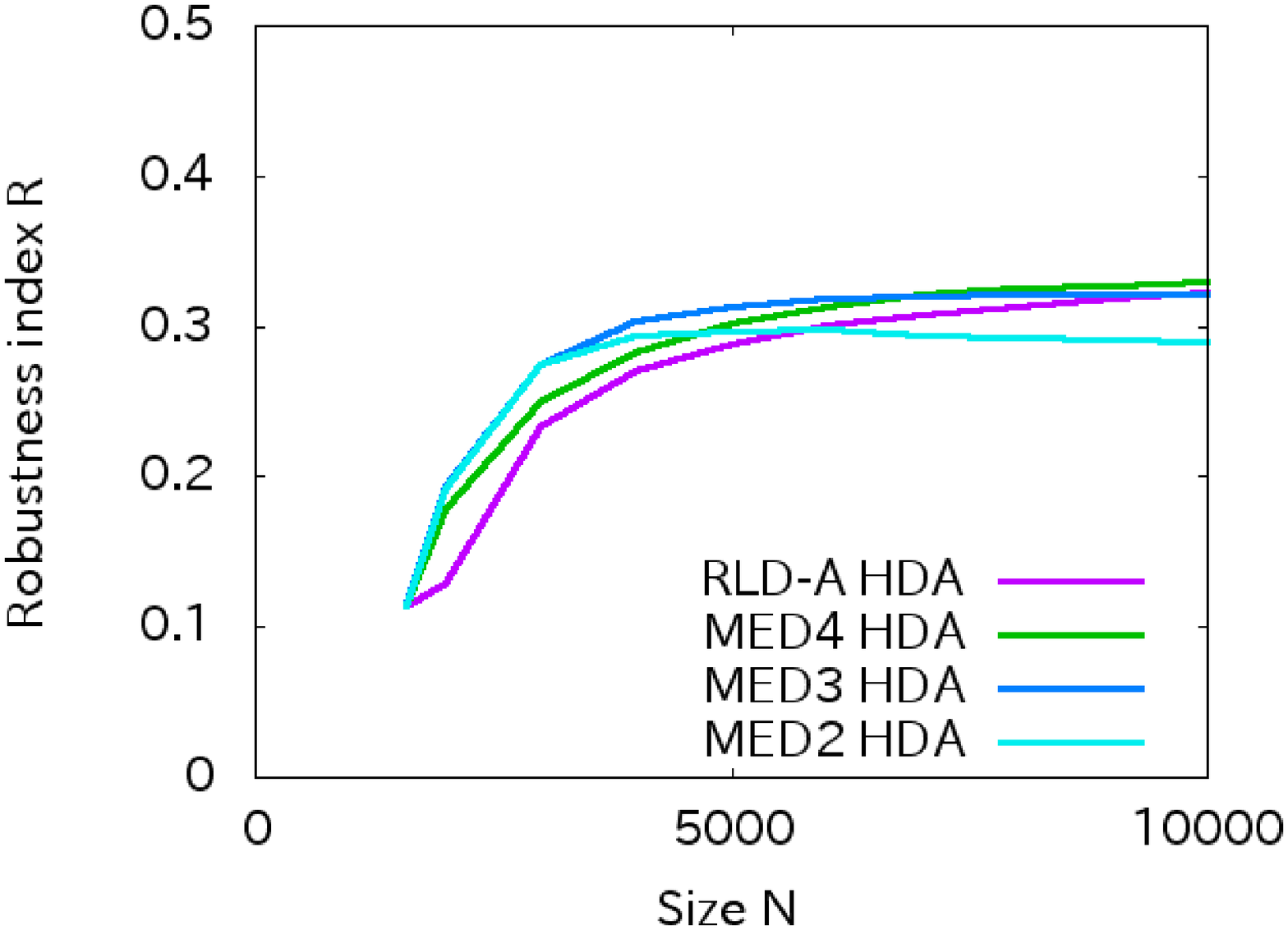}
\end{minipage}
\hfill
\begin{minipage}{.39\textwidth}
  \begin{center} c) Facebook for $m=10$ \end{center}
  \includegraphics[width=\textwidth]{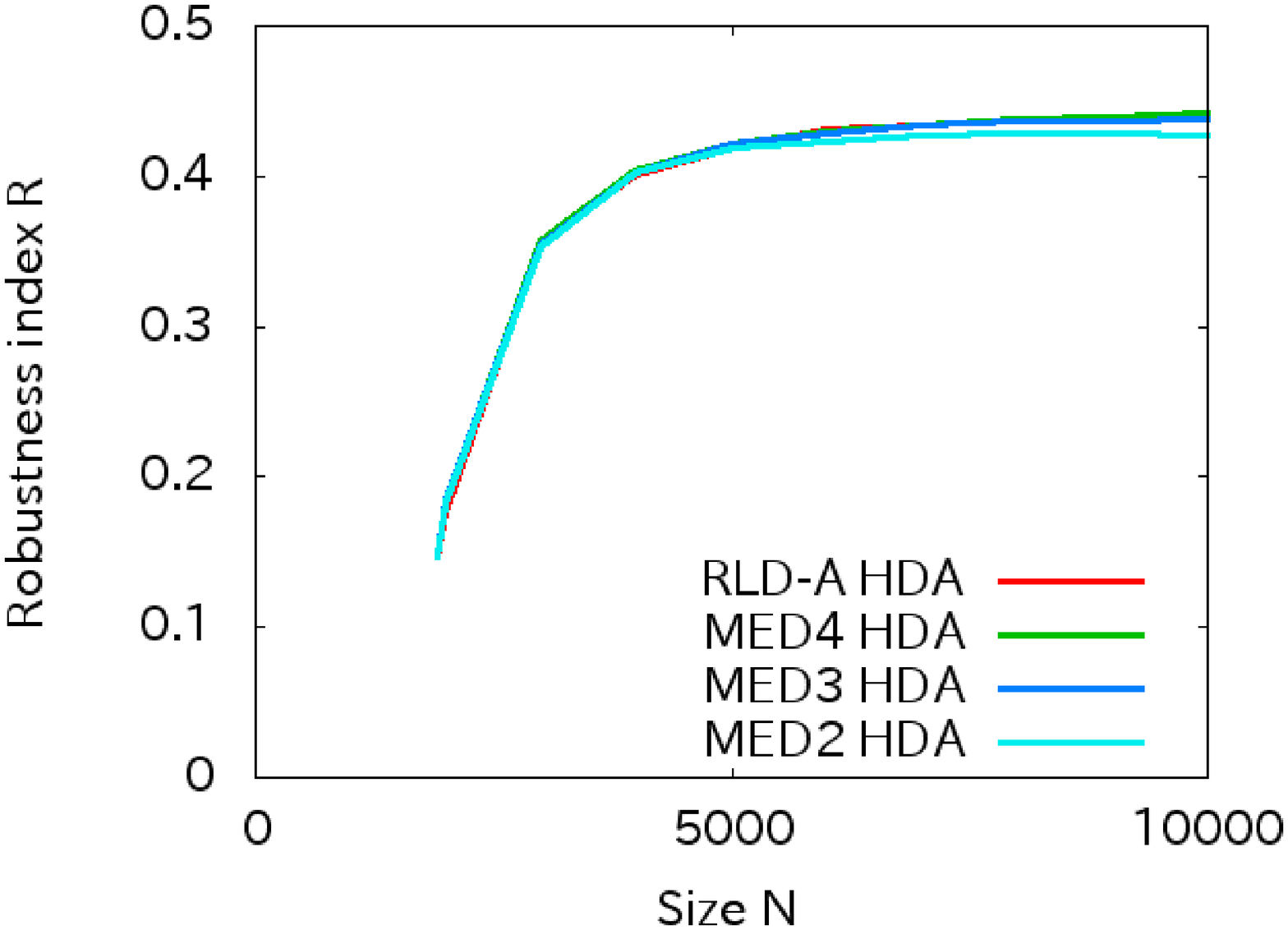}
\end{minipage}
\hfill
\begin{minipage}{.39\textwidth}
  \begin{center} d) US power grid for $m=4$ \end{center}
  \includegraphics[width=\textwidth]{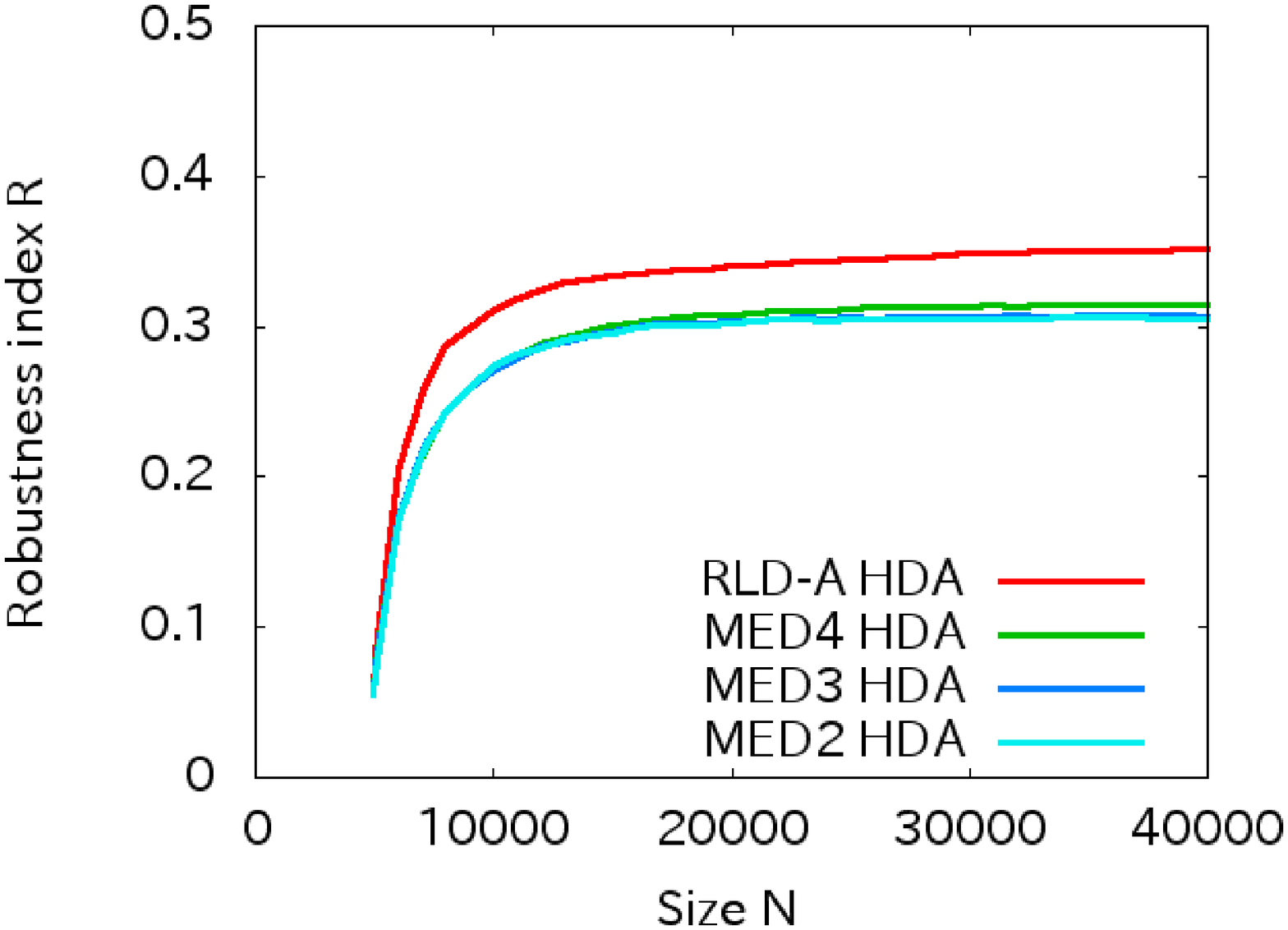}
\end{minipage}
\hfill
\begin{minipage}{.39\textwidth}
  \begin{center} e) US power grid for $m=12$ \end{center}
  \includegraphics[width=\textwidth]{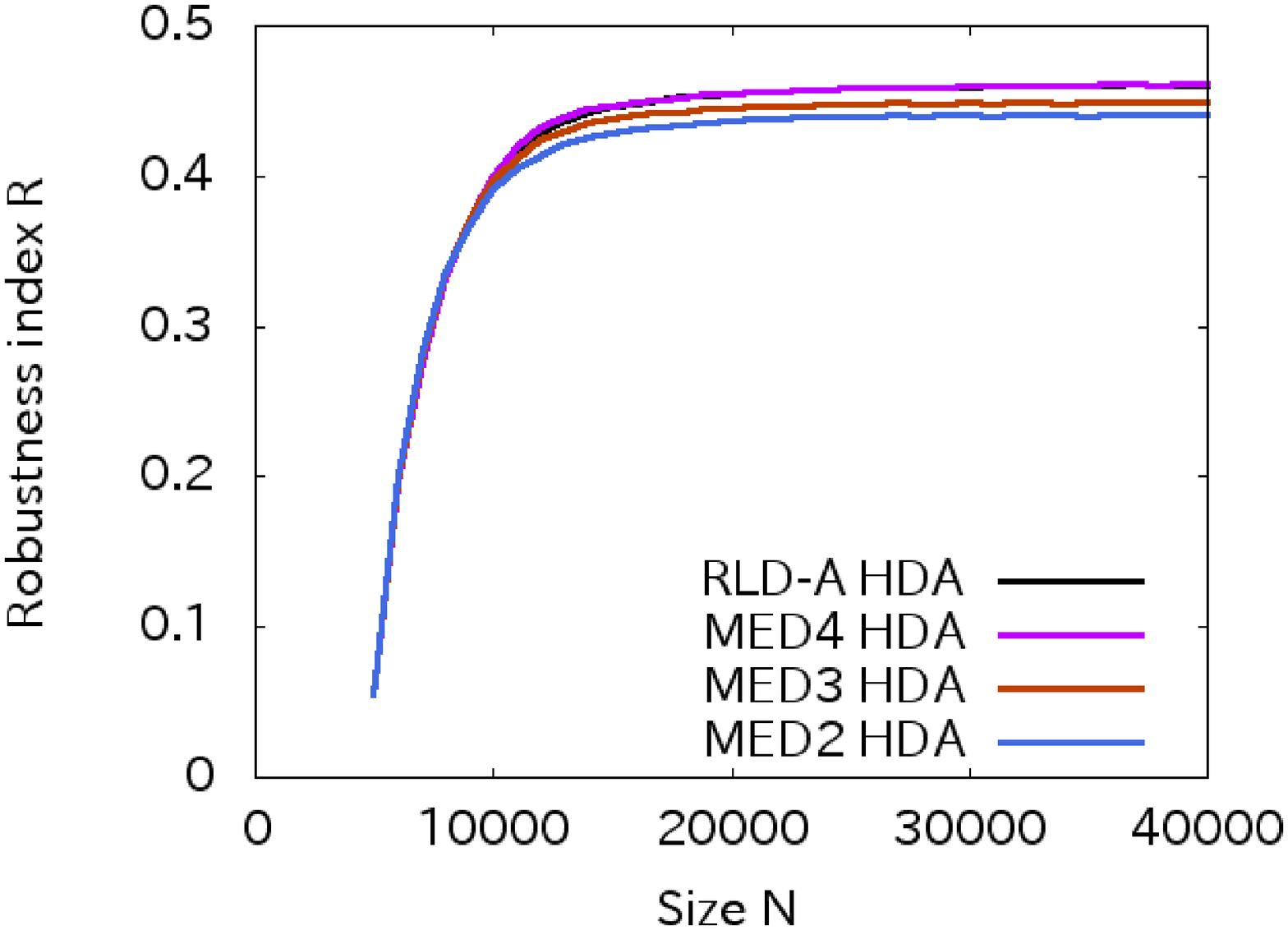}
\end{minipage}
\caption{Drastically improved robustness against HDA attacks 
in growing networks 
from the initial configuration of real networks. 
Robustness index $R$ vs $N$ in the growing networks
from the initial 
a) Facebook for $m=4$, 
b) US Airport Network for $m=4$, 
c) Facebook for $m=10$,
d) US power grid for $m=4$, and 
e) US power grid for $m=12$.
%In Figs. \ref{fig2}ab, 
%the robustness is improved with increasing $R$ by RLD-A 
%from vulnerable real networks. 
}
\label{fig7}
\end{figure}

\begin{figure}[htb]
\begin{minipage}{.39\textwidth}
  \begin{center} a) Facebook for $m=4$ \end{center}
  \includegraphics[width=\textwidth]{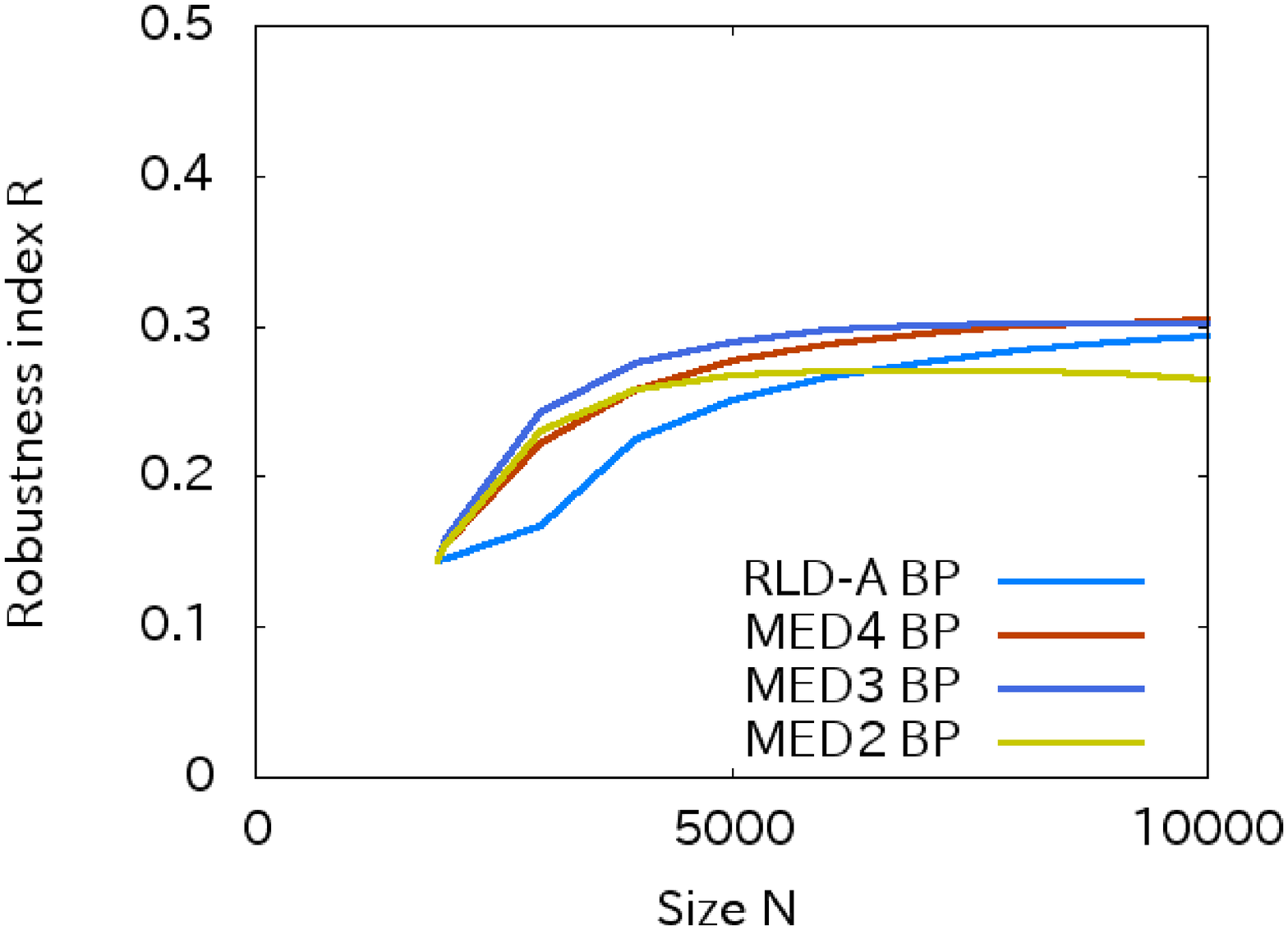}
\end{minipage}
\hfill
\begin{minipage}{.39\textwidth}
  \begin{center} b) US Airport Network for $m=4$ \end{center}
  \includegraphics[width=\textwidth]{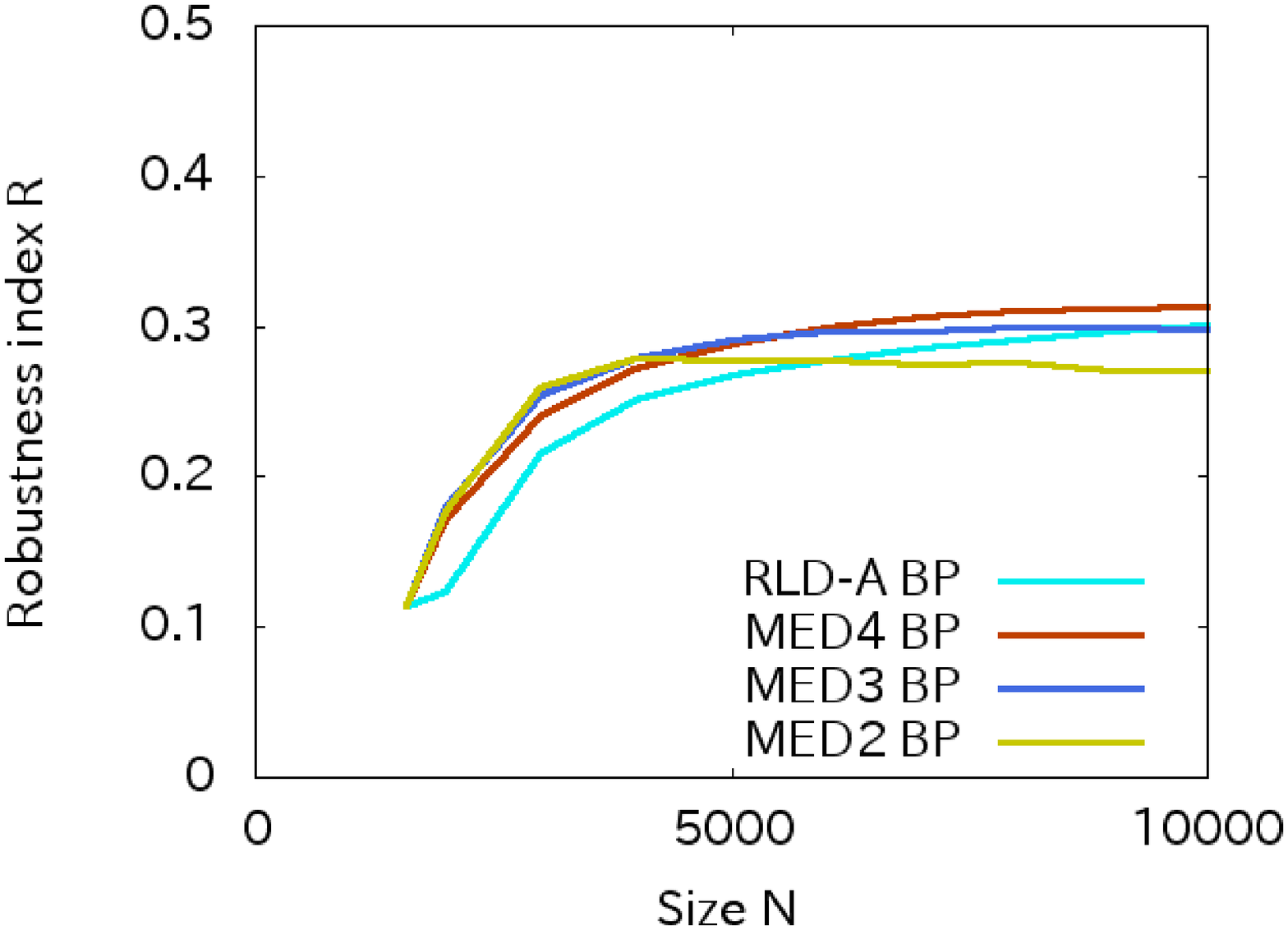}
\end{minipage}
\hfill
\begin{minipage}{.39\textwidth}
  \begin{center} c) Facebook for $m=10$ \end{center}
  \includegraphics[width=\textwidth]{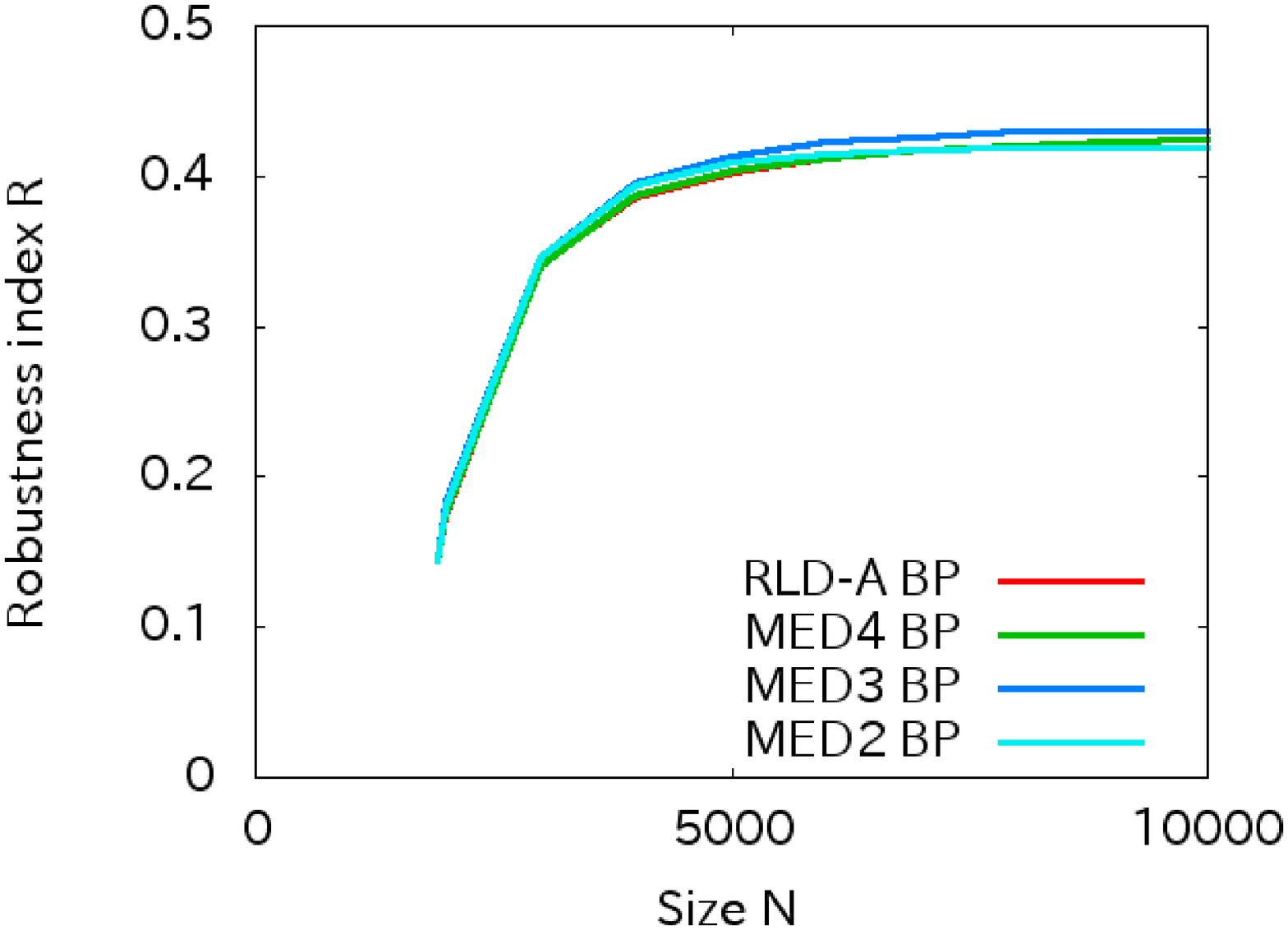}
\end{minipage}
\hfill
\begin{minipage}{.39\textwidth}
  \begin{center} d) US power grid for $m=4$ \end{center}
  \includegraphics[width=\textwidth]{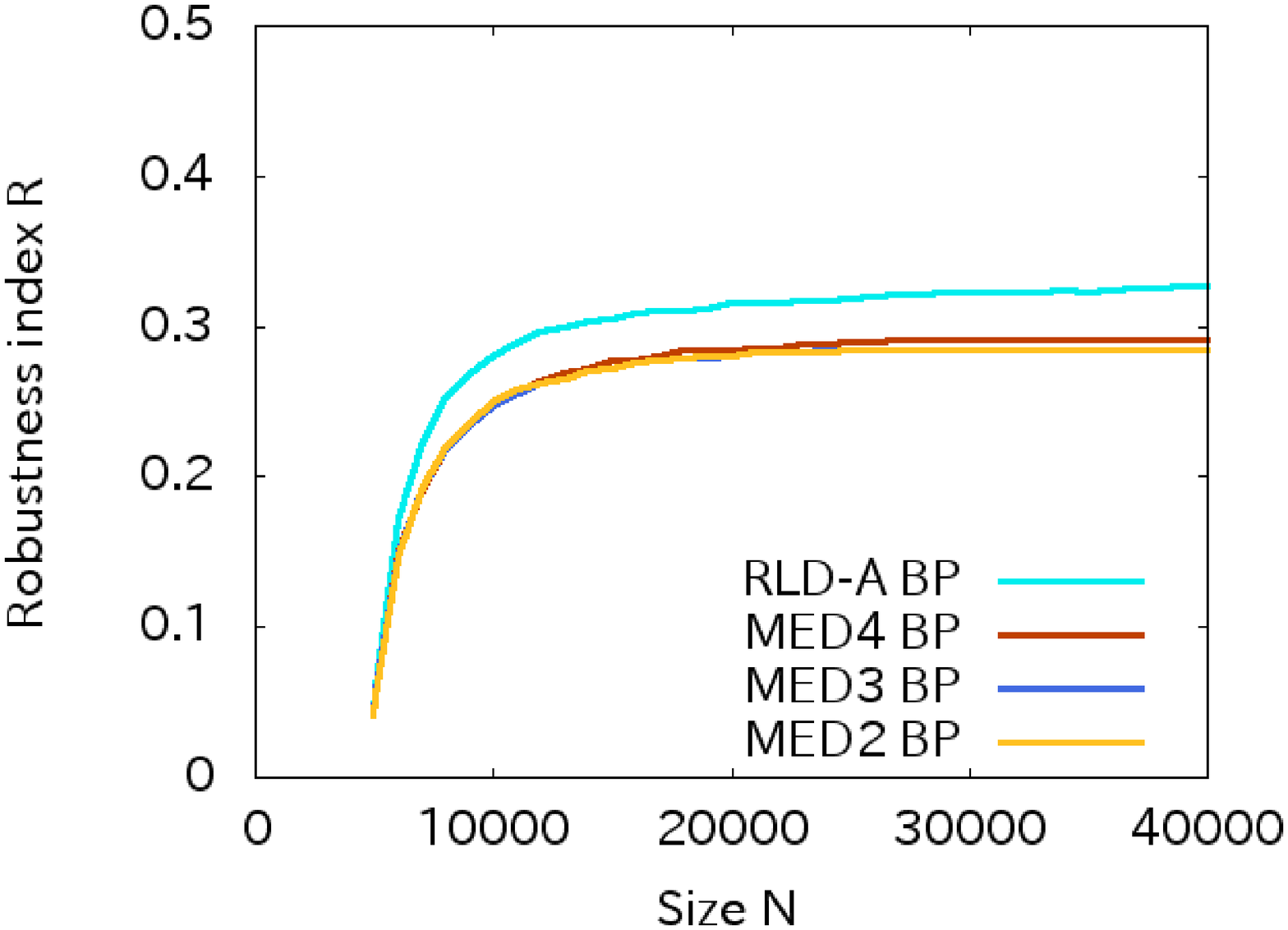}
\end{minipage}
\hfill
\begin{minipage}{.39\textwidth}
  \begin{center} e) US power grid for $m=12$ \end{center}
  \includegraphics[width=\textwidth]{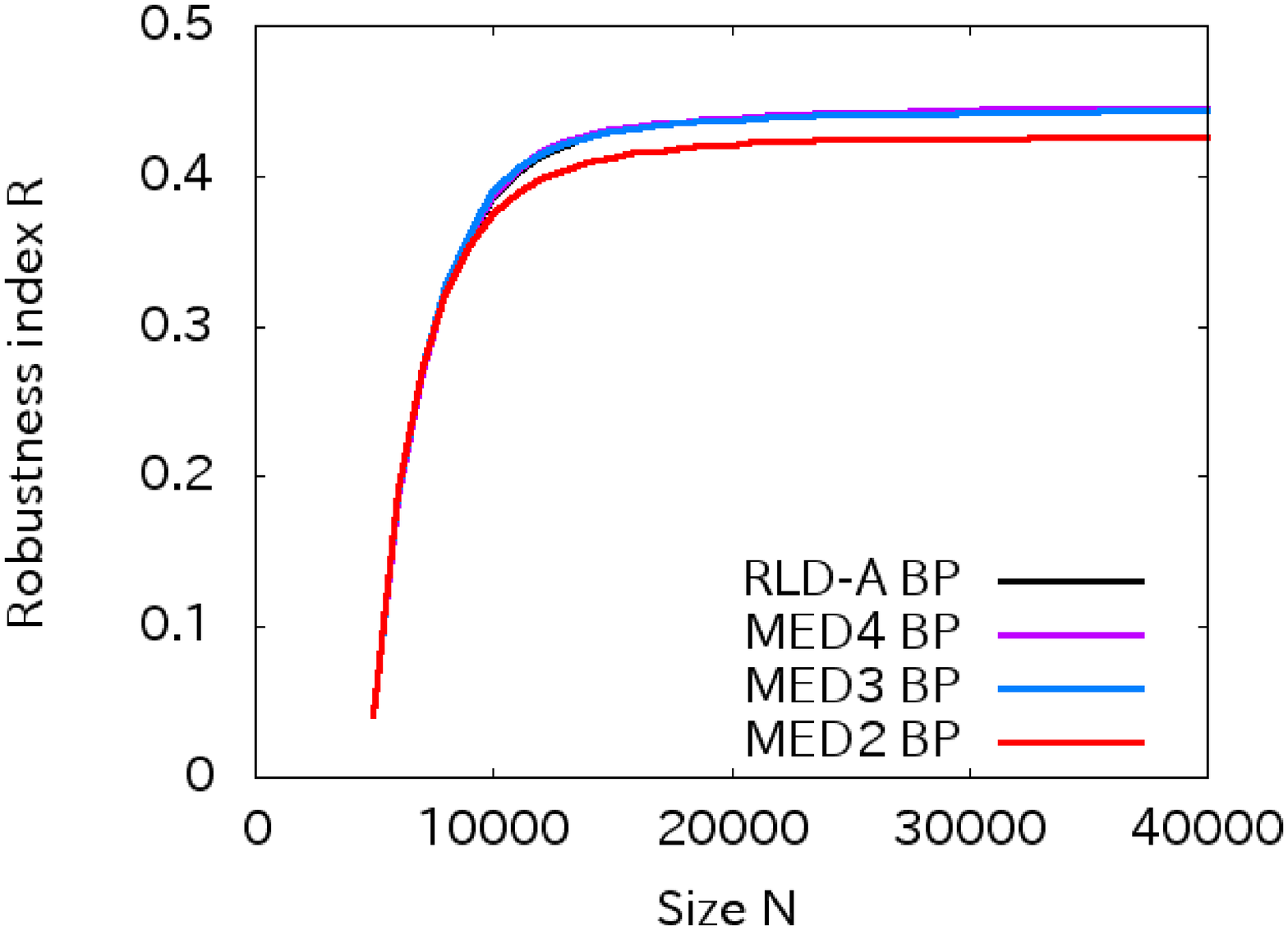}
\end{minipage}
\caption{Drastically improved robustness against BP attacks 
in growing networks from the initial configuration of real networks. 
Robustness index $R$ vs $N$ in the growing networks from the initial 
a) Facebook for $m=4$, 
b) US Airport Network for $m=4$, 
c) Facebook for $m=10$,
d) US power grid for $m=4$, 
and e) US power grid for $m=4$ and $m=12$.
}
\label{fig8}
\end{figure}

This network design in growth is different task from healing or 
recovering by rewirings e.g. between second neighbors 
\cite{Gallos15,Park16} 
in almost constant numbers of nodes and links 
for a damaged network by earthquakes or terrorist attacks, etc.
%Since this paper discuss incrementally growing networks with 
%an added node and attachment links per time step, 
%even if the attached nodes as rewirings are restricted between 
%second neighbors with respect to removed nodes, 
%only the pair of attachments is insufficient to determine 
%which existing node should be an anchor corresponded to our model's 
%new node, which parts should be rewired by just like RLD-A or MED, 
%how many rewirings are required at least 
%until reaching at what level of healing or recovering. 
%The spatial positions of nodes are also important to determine 
%them. 
%Many future issues can be considered for healing or recovering 
%by rewirings, they are beyond our current scope.
Because we focus on 
a structural change of network from almost uncorrelated SF 
to onion-like without hubs in the growth 
rather than topologically partial changes by rewirings. 
We also investigate dependence of the initial network structure 
not complete graph and the initial size $N_{0}$ 
on the robustness and degree-degree correlations 
for our growing method. 
Of course, some investments may be required for the growing network
in larger size than the initial real one, 
however the virtual test will give a prospective insight.

\begin{figure}[htb]
\begin{minipage}{.39\textwidth}
  \begin{center} a) Facebook for $m=4$ \end{center}
  \includegraphics[width=\textwidth]{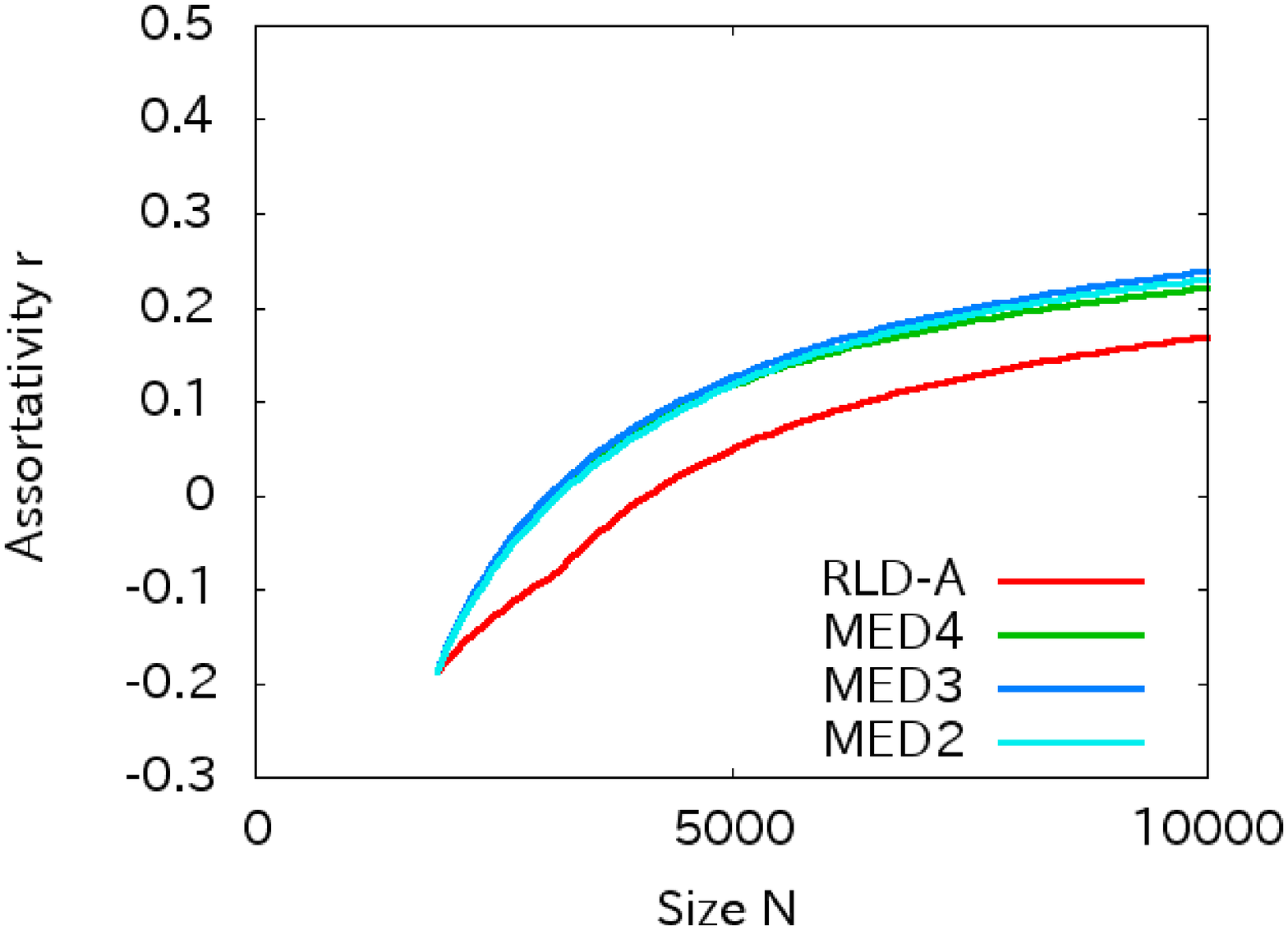}
\end{minipage}
\hfill
\begin{minipage}{.39\textwidth}
  \begin{center} b) US Airport Network for $m=4$ \end{center}
  \includegraphics[width=\textwidth]{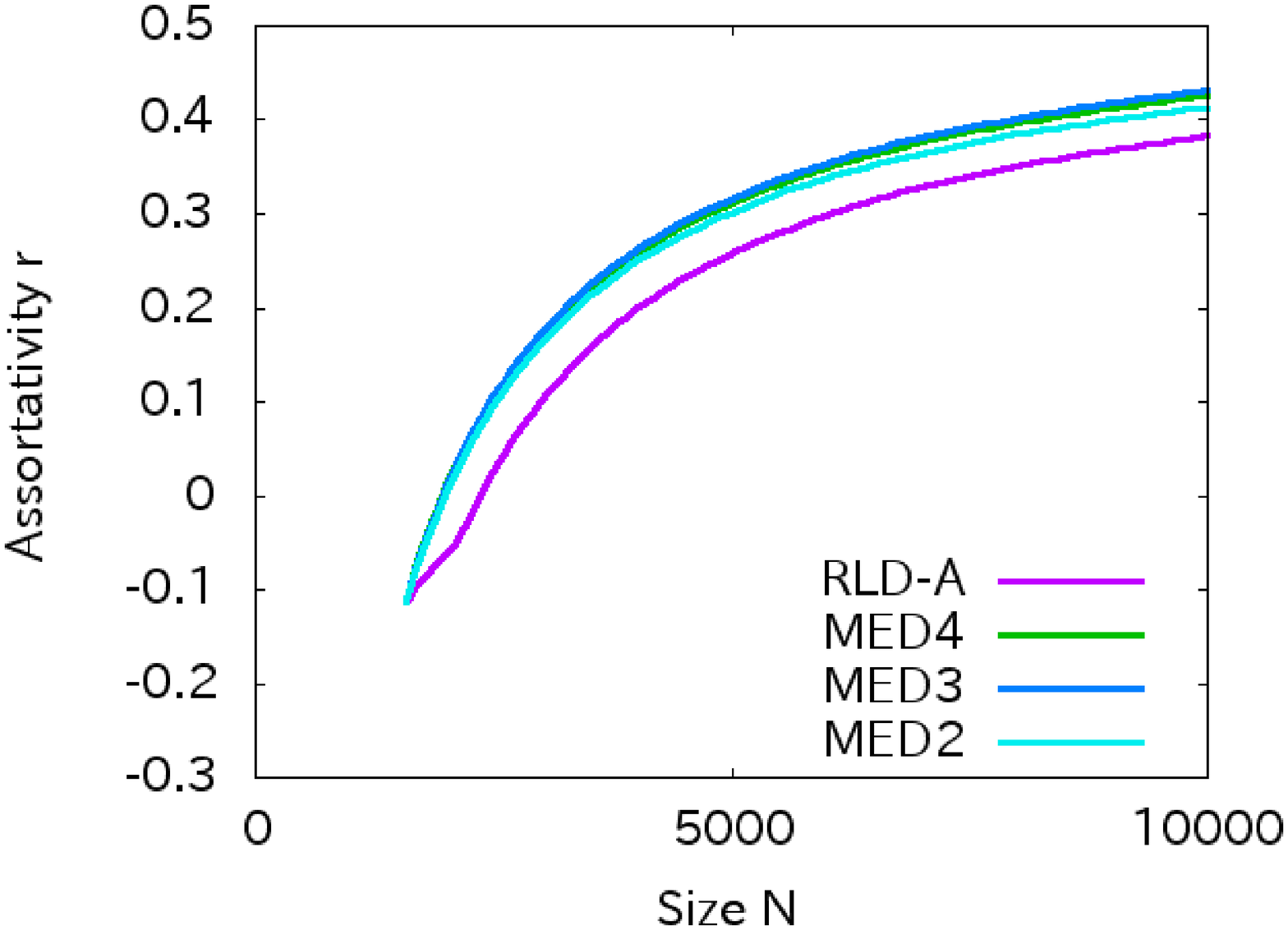}
\end{minipage}
\hfill
\begin{minipage}{.39\textwidth}
  \begin{center} c) Facebook for $m=10$ \end{center}
  \includegraphics[width=\textwidth]{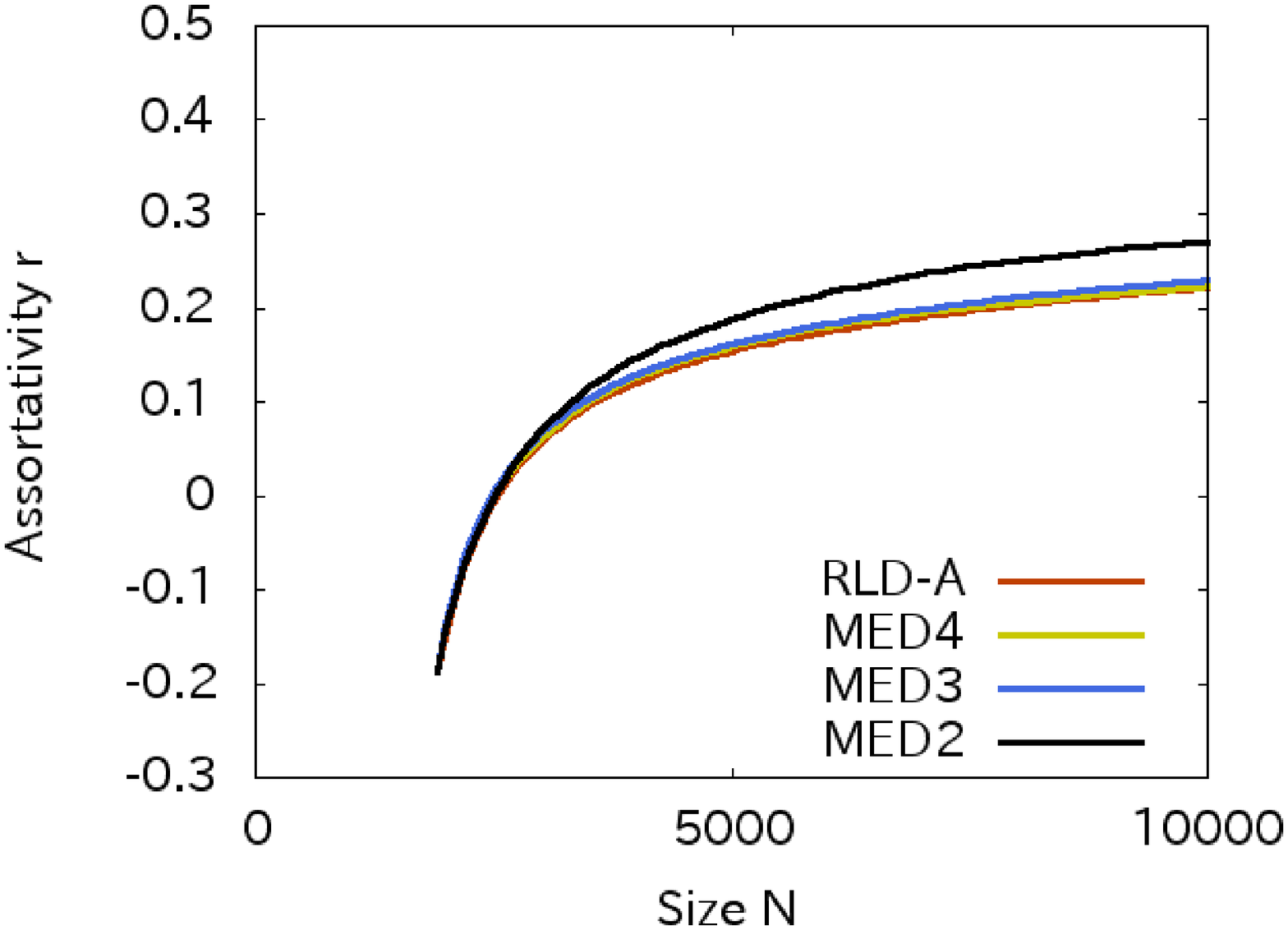}
\end{minipage}
\hfill
\begin{minipage}{.39\textwidth}
  \begin{center} d) US power grid for $m=4$ \end{center}
  \includegraphics[width=\textwidth]{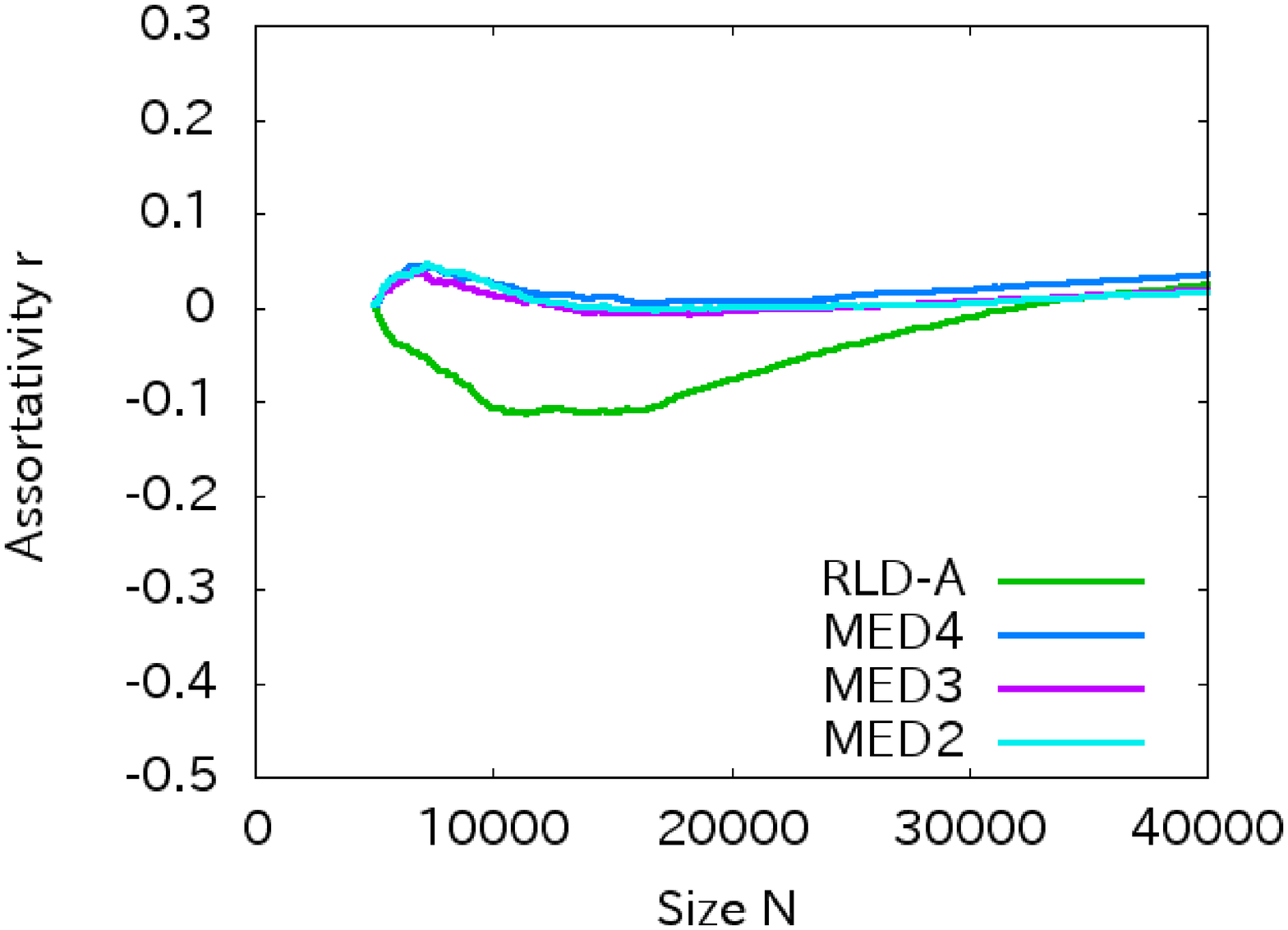}
\end{minipage}
\hfill
\begin{minipage}{.39\textwidth}
  \begin{center} e) US power grid for $m=12$ \end{center}
  \includegraphics[width=\textwidth]{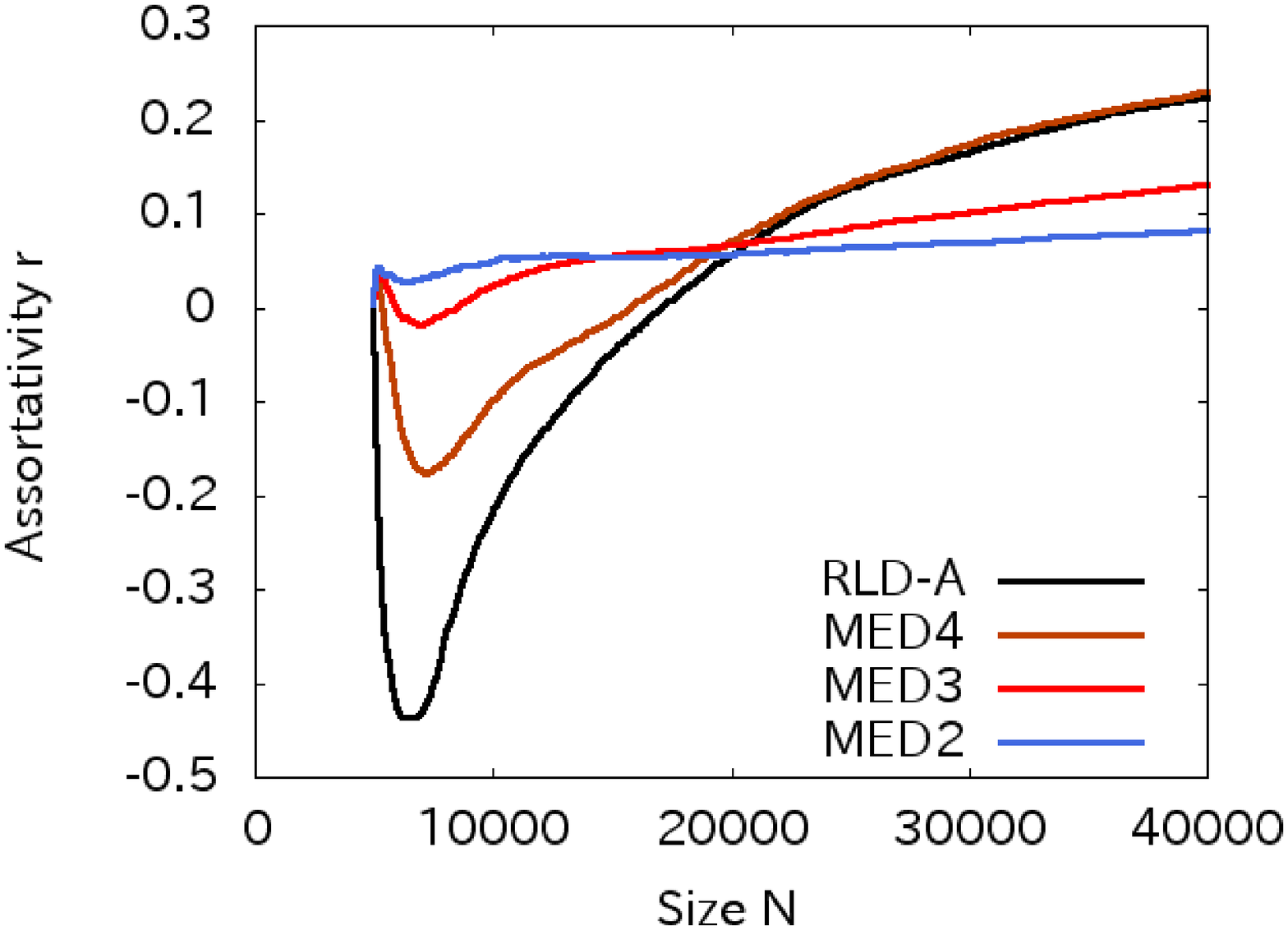}
\end{minipage}
\caption{Assortativity $r$ vs $N$ in growing networks from 
the initial 
a) Facebook for $m=4$, 
b) US Airport Network for $m=4$, 
c) Facebook for $m=10$, 
d) US Power Grid for $m=4$, and 
e) US Power Grid for $m=12$. 
%Until reaching positive degree-degree correlations, 
%the curves of $r$ behave differently from Fig. \ref{fig8}
}
\label{fig9}
\end{figure}

\begin{figure}[htb]
\begin{minipage}{.495\textwidth}
  \begin{center} a) Facebook and US Airport Network \end{center}
  \includegraphics[width=\textwidth]{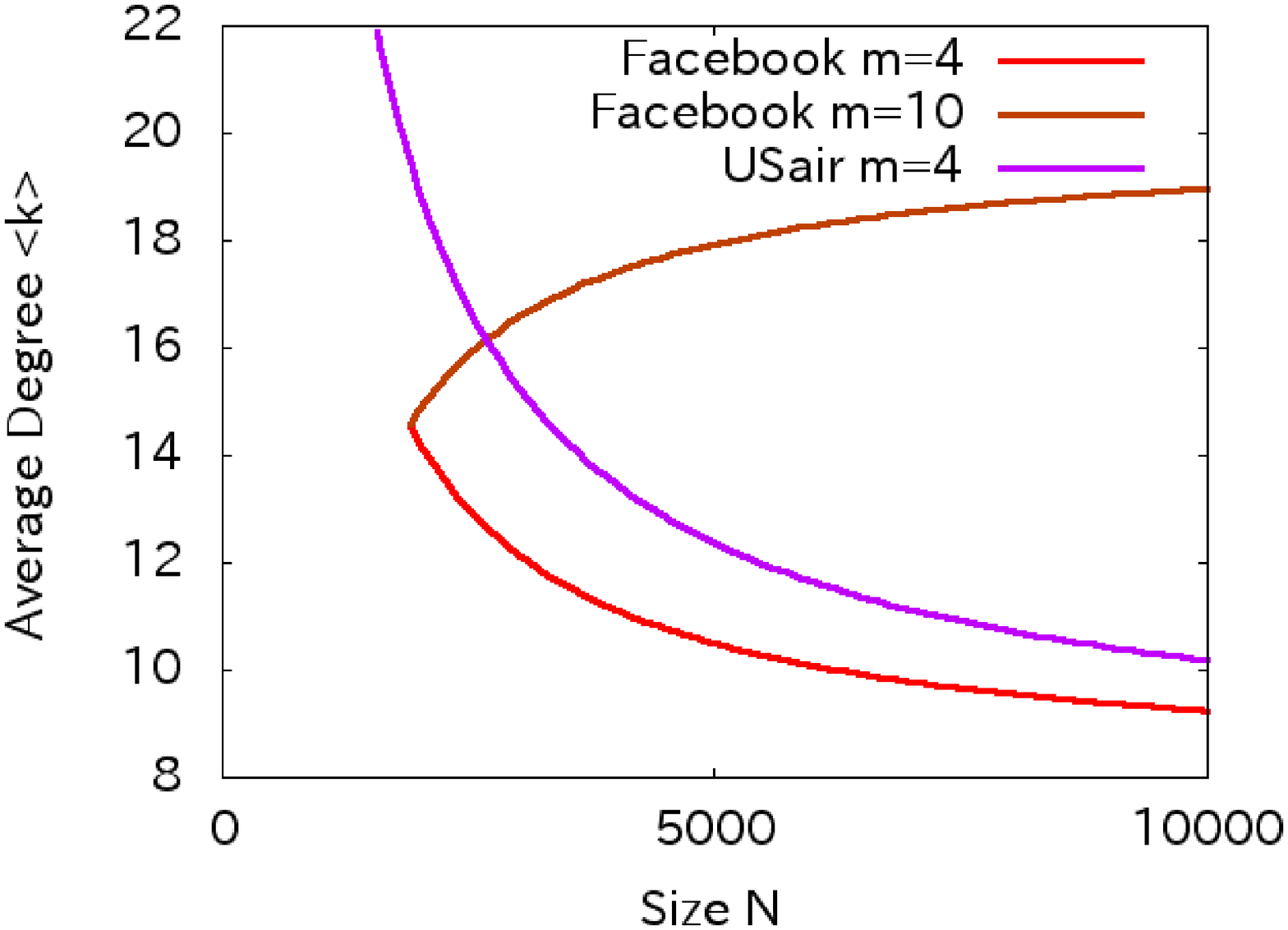}
\end{minipage}
\hfill
\begin{minipage}{.495\textwidth}
  \begin{center} b) US Power Grid \end{center}
  \includegraphics[width=\textwidth]{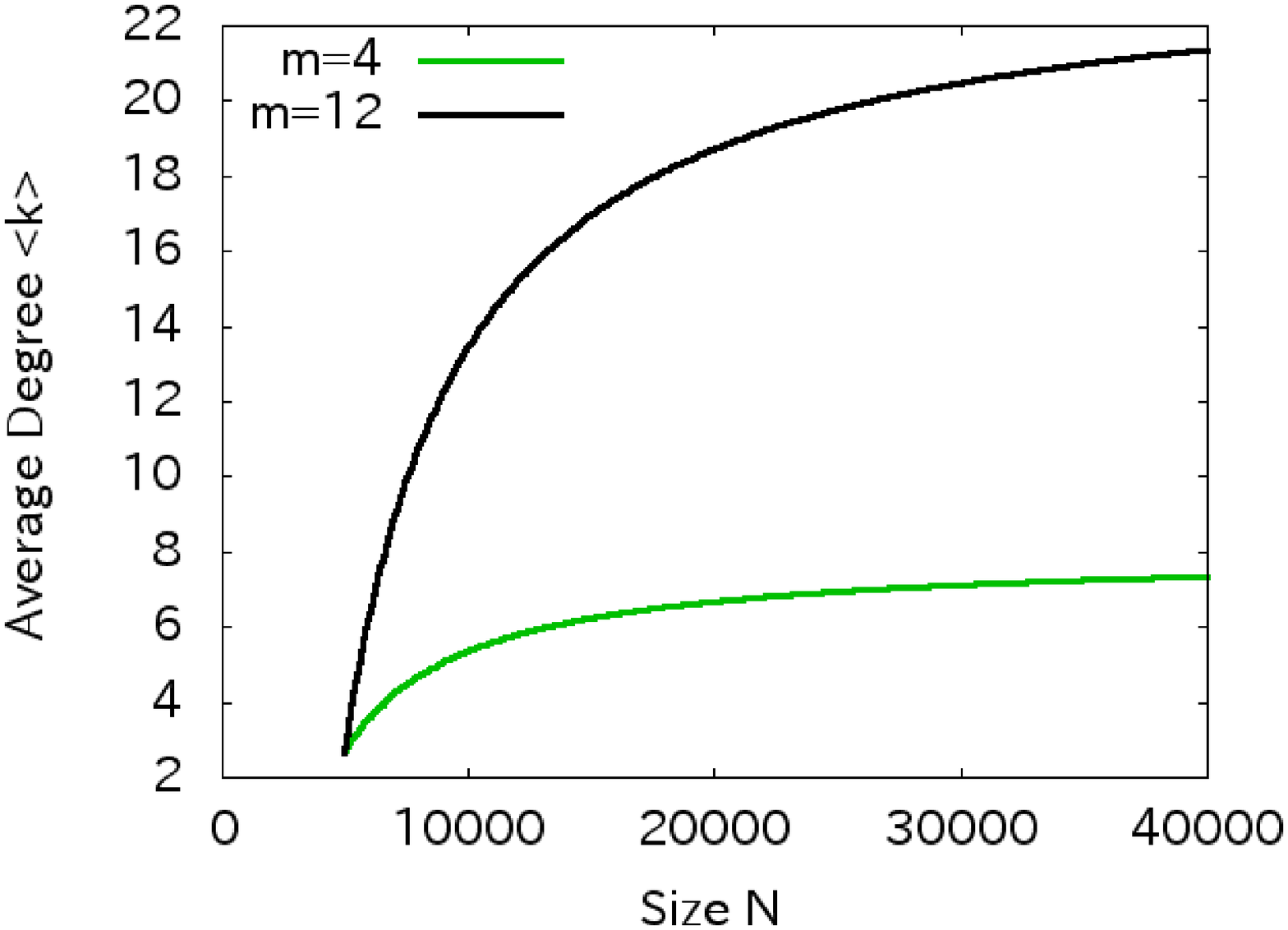}
\end{minipage}
\caption{Average degree $\langle k \rangle$ vs $N$ 
in the growing networks from the initial 
a) Facebook for $m=4$ and $m=10$ 
and US Airport Network for $m=4$, 
b) US Power Grid for $m=4$ and $m=12$.
}
\label{fig10}
\end{figure}

In the following, the cases of intermediately destructive 
CI$_{3}$ attacks and not very effective PLD-A are omitted to 
simplified the discussion.
Figures \ref{fig7}ab and \ref{fig8}ab 
show that 
high robustness against both HDA and BP attacks is obtained with 
increasing to $R > 0.3$ from 
$R \approx 0.1$ in initial vulnerable real networks of 
Facebook and USair.
In Figs. \ref{fig7}ce and \ref{fig8}ce, 
the cases of $m= 10$ and $m=12$ are investigated for checking 
the emergence of onion-like structure with high assortativity 
$r > 0.2$. 
We remark that some 
range-limited cases of MED in $\mu = 3$ intermediations 
(bluish lines) have higher $R$ than the cases of RLD-A with 
the attachments to the furthest nodes, 
but the effect is weak in the cases of MED in 
$\mu = 2$ intermediations.
If we do not insist onion-like networks, 
USpower can be already grown with high robustness before 
around $N \approx 10000$ of double size of the initial as shown 
in Figs. \ref{fig7}de and \ref{fig8}de. 
It suggests a possibility for incrementally growing 
other robust networks with $r < 0$ or $r \approx 0$ 
(see Fig. \ref{fig9}de) 
instead of onion-like networks. 
Figure \ref{fig9}abc shows the enhancing of degree-degree correlations 
in increasing values of $r$ in the growth of Facebook or USair. 
However, different behavior of late increasing after decreasing 
to negative correlations is obtained in USpower 
as shown in Fig. \ref{fig9}de. 
Some sort of structural change is occurred in the growth.
The reason may be from that 
USpower includes chain-like parts which induce large diameter, 
while Facebook and USair are compact with small diameter 
in Table \ref{table1}.
We note that the average path length is monotonously 
increasing to $4.13$ and $4.21$ at $N=10000$ in Facebook and USair, 
but decreasing to $4.97$ at $N=40000$ in USpower. 
These distributions of path lengths are bell-shaped with the 
peak around the average length for each size $N$.
In addition, 
the average path length in USpower is substantially greater 
than the number 
$\mu$ of intermediations at least in the early stage of the growth
as shown in Table \ref{table1}.

From Figs. \ref{fig7}, \ref{fig8}, \ref{fig9}, 
onion-like structure with both high $R$ and $r$ emerges by RLD-A or 
MED for $\mu \geq 3$ intermediations 
in the growing from Facebook for $m=10$, 
USair for $m=4$, and USpower for $m=12$. 
The growth to onion-like networks needs more steps 
as the initial size $N_{0}$ is larger in comparison of Facebook or 
USair and USpower. 
We remark that 
the effect of degree-degree correlations on the robustness 
works well in the early stage of the growth with structural change 
from real networks, 
however the robustness index $R$ is saturated in $N > 6000$ 
in Figs. \ref{fig7}abc and \ref{fig8}abc and 
in $N > 20000$ in Figs. \ref{fig7}de and \ref{fig8}de
for increasing assortativity $r$ with enhancing the correlations 
in Fig. \ref{fig9}. 
It is also interesting that the robustness 
in Figs. \ref{fig7} and \ref{fig8} is improved 
in spite of decreasing average degree $\langle k \rangle$ 
in the growing from Facebook and 
USair for $m=4$ (red and purple lines) as shown in Fig. \ref{fig10}a. 
Note that the average degree approaches to $2 \times m$ 
for $N \rightarrow \infty$ in Fig. \ref{fig10}.

\section{Conclusion}
We have proposed a second method for incrementally growing 
strongly robust onion-like networks self-organized by 
more natural and reasonable pair of attachments than the 
copying model \cite{Hayashi14,Hayashi16a}.
In addition, 
it becomes robust even in the early stage of the growth, 
and there is no huge hub whose largest degree is bounded. 
Since random attachments make an exponential degree 
distribution \cite{Barabasi99}, the random ones are dominant 
in the tail for high degrees, 
while another intermediated attachments mainly work for 
low degrees and the positive correlations among them. 
In a virtual test for the growing 
from real networks with extreme vulnerability, 
we have shown that the proposed growing networks have 
reformable robustness 
to be future prospective infrastructures. 
It is also expected that 
the range-limited intermediations in a few hops 
reduce the Euclidean distances of links embedded on a space.

We emphasize the emergence of robust onion-like networks 
in enhancing moderately long loops 
by range-limited MED in a few hops 
without both 
degrading efficiency of paths and large connection costs or efforts. 
We should remember that the coexistence of robustness and 
efficiency has not been realized in many real networks, and 
the threat against attacks \cite{Makse15,Zhou16}
is never decreased rather increased more and more, 
unless the dependence on selfish preferential attachment \cite{Barabasi99} 
is changed by ourselves.
Therefore, our study suggests that 
we should discontinue the dependence on selfish rule 
and develop the potential of distant connections for the half of links, 
which may mean necessary investment for highly reliable 
connectivity in our network infrastructures 
even against intelligent attacks.

\section*{Acknowledgments}
%The author would like to thank 
The author express appreciation to anonymous reviewers and the web site 
noted in footnote 
for their valuable comments and using dataset of real networks.
This research was supported in part by 
JSPS KAKENHI Grant Number JP.17H01729.
%a Grant-in-Aid for Scientific Research in Japan, No. JP25330100.

\section*{Appendix}
We briefly summarize the theoretical background of CI and BP.

For CI according to \cite{Makse15}, 
we consider the optimal immunization 
to prevent disease or information spreading. 
This is mapped onto the breakdown problem as 
minimizing the size of GC of a network. 
In the GC after removing a fraction $q$ of nodes, 
there are two states represented by the quantity 
$n_{i} = 1$ or $n_{i} = 0$: node $i$ exists or not. 
The fraction $q$ is denoted by 
$q = 1 - \frac{1}{N} \sum_{i=1}^{N} n_{i}$.

The probability $\nu_{i \rightarrow j}$ for information spreading 
is computed in self-consistency through the following 
message passing equations 
\begin{equation}
  \nu_{i \rightarrow j} = n_{i} \left[ 1 -
    \Pi_{k \in \partial i \backslash j} (1 - \nu_{k \rightarrow i}) \right],
\label{eq_CI_MP}
\end{equation}
where $\partial i$ denotes node $i$'s set of connecting neighbor nodes, 
and $\partial i \backslash j$ is the node subset obtained by removing 
node $j$ from $\partial i$. 
Since the stability condition at the fixed point of origin 
for the iterated map of Eq. (\ref{eq_CI_MP}) 
is given by that the largest eigenvalue $\lambda(n;q)$ of 
the Jacobian matrix 
\[
  \left. 
  \frac{\partial \nu_{i \rightarrow j}}{\partial \nu_{k \rightarrow l}}
  \right|_{\nu_{i \rightarrow j}=0}
  = n_{i} B_{k \rightarrow l, i \rightarrow j}
\]
is less than one, to find the minimum set of immunization nodes 
is equivalent to select the minimum set of removed nodes until 
satisfying $\lambda(n;q) = 1$. 
Here, $B_{k \rightarrow l, i \rightarrow j}$ is an element of 
$2M \times 2M$ non-backtracking (NB) matrix \cite{Hashimoto89} 
for the network. 
At the critical $\lambda(n;q_{c}) = 1$, the GC is broken.

Moreover, according to Power Method, we have 
\begin{equation}
  \lambda(n;q) = \lim_{l \rightarrow \infty}
  \left[ \frac{| {\bf w}_{l}(n) |}{| {\bf w}_{0} |} \right]^{1/l},
\label{eq_lambda}
\end{equation}
${\bf w}_{l}(n)$ denotes the vector at the $l$-th iterations by 
multiplying the Jacobian matrix
for an initial arbitrary nonzero vector ${\bf w}_{0}$. 
The elements of $2 M$ dimensional vectors ${\bf w}_{l}(n)$ 
and ${\bf w}_{0}$ are for bidirectional 
$i \rightarrow j$ and $j \rightarrow i$ links. 
The denominator $| {\bf w}_{0} |$ 
in the right-hand side of Eq. (\ref{eq_lambda}) is constant 
and independent of the set $\{ n_{i} \}$. 
The numerator 
is approximated by the following expression corresponding to 
$2 l$-body problem.
\begin{equation}
  | {\bf w}_{l}(n) |^{2} \approx \sum_{i=1}^{N} (k_{i} - 1) 
  \sum_{j \in \partial Ball(i,2l-1)} \left( \Pi_{k \in P_{2l-1}(i,j)} 
  n_{k} \right) (k_{j} - 1), 
\label{eq_2l-body}
\end{equation}
where $P_{2l-1}(i,j)$ is the set of nodes belonging to the 
shortest path of length $2l-1$ connecting $i$ and $j$ nodes, 
$\partial Ball(i,l)$ denotes the set of nodes on the frontier 
of the ball with radius $l$ hops from node $i$, 
$k_{i}$ and $k_{j}$ denotes the degrees of nodes $i$ and $j$. 
We remark the factor $\Pi_{k} n_{k} = 1$ in the parenthesis 
in the right-hand side of Eq. (\ref{eq_2l-body}) 
when all nodes on the shortest path are not absence, 
then the path is connected.
The above 
approximation can be extended to the case of even length path 
to $\partial Ball(i, 2l)$.
Thus, for $i = 1, 2, \ldots, N$, 
the expression in the right-hand side of Eq. (\ref{eq_2l-body}) gives 
\begin{equation}
  CI_{l}(i) \stackrel{\rm def}{=} (k_{i} - 1) 
  \sum_{j \in \partial Ball(i,l)} (k_{j} - 1). \label{eq_CI}
\end{equation}

This approximation of CI in $l$ hops is categorized 
in range-limited approach. 
Thus, the scalable algorithm for calculating $CI_{l}(i)$ 
is based on the minimization of the energy of a many-body system, 
which is equivalent 
to find the principal part of connectivity as 
the minimal set of nodes that minimize the largest eigenvalue 
$\lambda(n, q)$.
The critical transition of the eigenvalue of the NB matrix 
from one to zero is 
caused as the network consists of a single loop 
is changed to a tree by a node removal eventually. 
Once it becomes a tree, fragmentation to large components 
occurs by any node removal. 
When a network includes more than one loops, the eigenvalue is 
greater than one. 
Thus, it has been pointed out that 
{\it the best attack strategy is to destroy the loops}
\cite{Makse15}(Supplementary Information). 
Note that the NB matrix is intrinsic for enumerating 
the number of loops on a length basis 
in Zeta function of graphs \cite{Hashimoto89}.

\vspace{3mm}
For BP according to \cite{Zhou13}, 
it is assumed that 
nodes $j \in \partial i$ are mutually independent of each other 
when node $i$ is removed. 
Such approximated tree-like graph is called as cavity graph.
We consider the marginal probability $q^{A_{i}}_{i}$ for 
the state $A_{i}$ of node $i$.
Since $A_{i}$ represent the index of root node of $i$, 
it is influenced by the neighboring nodes in the cavity graph 
after removing node $i$ denoted by $\backslash i$.
Based on the product of independent marginal probability 
$q^{A_{j}}_{j \rightarrow i}$ for the state $A_{j}$, 
we consider the joint probability 
\[
  {\cal P}_{\backslash i}({A_{j}: j \in \partial i}) 
  \approx \Pi_{j \in \partial i} q^{A_{j}}_{j \rightarrow i}.
\]
In the cavity graph, 
if all nodes $j \in \partial i$ are either empty ($A_{j} = 0$)
or roots ($A_{j} = j$), 
the added node $i$ can be a root ($A_{i} = i$).
There are the following exclusive states.
\begin{enumerate}
  \item $A_{i} = 0$: $i$ is empty (removed). 
    Since $i$ is unnecessary as a root, it belongs to FVS.
  \item $A_{i} = i$: $i$ becomes its own root.\\
    The state $A_{j} = j$ of $j \in \partial i$ 
    is changeable to $A_{j} = i$ when node $i$ is added.
  \item $A_{i} = k$: one node $k \in \partial i$ becomes the root
    of $i$ when it is added, 
    if $k$ is occupied and all other $j \in \partial i$ 
    are either empty or roots.
\end{enumerate}
The corresponding probabilities to the above states are 
\begin{equation}
  q^{0}_{i} \stackrel{\rm def}{=} \frac{1}{z_{i}}, \label{eq_BP1}
\end{equation}
\[
  q^{i}_{i} \stackrel{\rm def}{=} 
  \frac{e^{x} \Pi_{j \in\partial i(t)} 
  \left[ q^{0}_{j \rightarrow i} + q^{j}_{j \rightarrow i} \right]}{z_{i}},
\]
\[
 q^{k}_{i} \stackrel{\rm def}{=} 
 \frac{e^{x} \frac{(1 - q^{0}_{k \rightarrow i})}{
        q^{0}_{k \rightarrow i} + q^{k}_{k \rightarrow i}} 
  \Pi_{j \in\partial i(t)} 
  \left[ q^{0}_{j \rightarrow i} + q^{j}_{j \rightarrow i} \right]}{z_{i}},
\]
\begin{equation}
  q^{0}_{i \rightarrow j} = \frac{1}{z_{i \rightarrow j}(t)}, 
\end{equation}
\begin{equation}
  q^{i}_{i \rightarrow j} = 
  \frac{e^{x} \Pi_{k \in \partial i(t)\backslash j} 
    \left[ q^{0}_{k \rightarrow i} + q^{k}_{k \rightarrow i} \right]
  }{z_{i \rightarrow j}(t)}, 
\end{equation}
where $\partial i(t)$ denotes node $i$'s set of connecting neighbor nodes 
at time $t$, 
and $x > 0$ is a parameter of inverse temperature. 
We have the normalization constant 
\begin{equation}
  z_{i} \stackrel{\rm def}{=} 
  1 + e^{x} \left[ 1 + \sum_{k \in \partial i(t)} 
      \frac{1 - q^{0}_{k \rightarrow i}}{
        q^{0}_{k \rightarrow i} + q^{k}_{k \rightarrow i}} \right]
  \Pi_{j \in\partial i(t)} 
  \left[ q^{0}_{j \rightarrow i} + q^{j}_{j \rightarrow i} \right],
  \label{eq_BP4}
\end{equation}
\begin{equation}
  z_{i \rightarrow j}(t) \stackrel{\rm def}{=} 
  1 + e^{x} \Pi_{k \in \partial i(t)\backslash j} 
  \left[ q^{0}_{k \rightarrow i} + q^{k}_{k \rightarrow i} \right]
  \times \left[ 1 + \sum_{l \in \partial i(t)\backslash j} 
    \frac{1 - q^{0}_{l \rightarrow i}}{
      q^{0}_{l \rightarrow i} + q^{l}_{l \rightarrow i}} \right], 
  \label{eq_BP5}
\end{equation}
to be satisfied for any $i$ and $i \rightarrow j$ as 
\[
  q^{0}_{i} + q^{i}_{i} + \sum_{k \in \partial i} q^{k}_{i} = 1,
\]
\[
  q^{0}_{i \rightarrow j} + q^{i}_{i \rightarrow j} 
  + \sum_{k \in \partial i} q^{k}_{i \rightarrow j} = 1.
\]

In BP attacks with sudden breakdown \cite{Zhou16}
which gives severer damages than CI and hub attacks, 
the node deletion process also focus on the destroy of loops, 
since {\it the FVS of graph $G$ 
is a subset of nodes such that 
if all the nodes of this set and the attached links are removed 
from $G$ the remaining graph will have no loops} \cite{Zhou13}. 
A node with the highest $q^{0}_{i}$ is chosen as the removed target
at each time step $t$ that consists of a number of rounds by 
the updating calculations of Eqs. (\ref{eq_BP1})-(\ref{eq_BP5}) 
in order of random permutation of nodes $1 \sim N$.

On the other hand in computer science, 
to find a suitable node for removing and inserting into FVS, 
the largest node of its degree $deg(v)$ minus the number $comp(G-v)$
of connecting components formed by removing the node $v$ 
is recursively chosen in an approximation algorithm \cite{Vazirani01}.
When several links connect to a component, pairs of 
these links form loops through $v$.
Then the number of them 
(precisely the number $-1$) is not decreased from its degree. 
The most subtracted case is only one link connects to a component, 
then $deg(v) - comp(G-v)$ becomes $0$.
Thus, $deg(v) - comp(G-v)$ is considered as 
a characteristic index to delete loops as many as possible.
However, this heuristic approximation method requires large 
computation.

\end{document}